\documentclass[twocolumn]{aastex62}  
\usepackage{xcolor, bm, amsmath, enumitem}
\usepackage[hang]{footmisc}
\usepackage{mfirstuc}
\usepackage{array}
\newcolumntype{H}{>{\setbox0=\hbox\bgroup}c<{\egroup}@{}}

\newcommand{\rutgers}{Rutgers University, Department of Physics and Astronomy, 136 Frelinghuysen Road, Piscataway, NJ 08854, USA}
\newcommand{\utaustin}{Department of Astronomy, The University of Texas at Austin, 2515 Speedway, Stop C1400, Austin, TX 78712-1205, USA}

\newcommand{\trans}[3]{\capitalisewords{#1}\,\textsc{#2}\,#3\,\AA{}}
\newcommand{\spec}[2]{\capitalisewords{#1}\,\textsc{#2}}
\newcommand{\hst}{\textit{HST}}
\newcommand{\msun}{$M_\odot$}
\newcommand{\mdot}{$\dot{M}_\star$}
\newcommand{\vsini}{$v \,\mathrm{sin}(i)$}
\newcommand{\vinf}{$v_\infty$}
\newcommand{\vrad}{$v_\mathrm{radial}$}
\newcommand{\vsys}{$v_\mathrm{sys}$}
\newcommand{\vedge}{$v_\mathrm{edge}$}
\newcommand{\hii}{H\,\textsc{ii}}

\newcommand{\heii}{He\,\textsc{ii}}
\newcommand{\teff}{$T_\mathrm{eff}$}
\newcommand{\logg}{$\log(g)$}

\newcommand{\lstar}{$L_\star$}
\newcommand{\mstar}{$M_\star$}
\newcommand{\rstar}{$R_\star$}
\newcommand{\zsun}{$Z_\odot$}

\newcommand{\chisq}{$\chi^2$}
\newcommand{\fracneb}{$f^\mathrm{nebular}_\mathrm{FUV}$}

\newcommand{\tlusty}{\textsc{tlusty}}
\newcommand{\dolphot}{\textsc{dolphot}}
\newcommand{\beast}{\textsc{beast}}
\newcommand{\cmfgen}{\textsc{cmfgen}}
\newcommand{\parsec}{\textsc{parsec}}

\shorttitle{O-Dwarf Stars at Extremely Low Metallicity}
\shortauthors{Telford et al.}

\begin{document}

\title{Far-Ultraviolet Spectra of Main-Sequence O Stars at Extremely Low Metallicity}
\author[0000-0003-4122-7749]{O. Grace Telford}
\affiliation{\rutgers}
\email{grace.telford@rutgers.edu}
\author[0000-0002-0302-2577]{John Chisholm}
\affiliation{\utaustin}
\author[0000-0001-5538-2614]{Kristen B. W. McQuinn}
\affiliation{\rutgers}
\author[0000-0002-4153-053X]{Danielle A. Berg}
\affiliation{\utaustin}


\begin{abstract}

Metal-poor massive stars dominate the light we observe from star-forming dwarf galaxies and may have produced the bulk of energetic photons that reionized the universe at high redshift.
Yet, the rarity of observations of individual O stars below the 20\% solar metallicity (\zsun{}) of the Small Magellanic Cloud (SMC) hampers our ability to model the ionizing fluxes of metal-poor stellar populations.
We present new \textit{Hubble Space Telescope} far-ultraviolet (FUV) spectra of three O-dwarf stars in the galaxies Leo~P (3\%\,\zsun{}), Sextans~A (6\%\,\zsun{}), and WLM (14\%\,\zsun{}). 
We quantify equivalent widths of photospheric metal lines and strengths of wind-sensitive features, confirming that both correlate with metallicity.
We infer the stars' fundamental properties by modeling their FUV through near-infrared spectral energy distributions and identify stars in the SMC with similar properties to each of our targets.
Comparing to the FUV spectra of the SMC analogs suggests that (1) the star in WLM has an SMC-like metallicity,  and (2) the most metal-poor star in Leo~P is driving a much weaker stellar wind than its SMC counterparts.
We measure projected rotation speeds and find that the two most metal-poor stars have high \vsini{}\,$\geq$\,290\,km\,s$^{-1}$, and estimate just a $3-6\%$ probability of finding two fast rotators if the metal-poor stars are drawn from the same \vsini{} distribution observed for O dwarfs in the SMC.
These observations suggest that models should include the impact of rotation and weak winds on ionizing flux to accurately interpret observations of metal-poor galaxies in both the near and distant universe.
\end{abstract}


\section{Introduction\label{sec:intro}}

\subsection{Uncertainties in Modeling Metal-Poor Massive Stellar Populations}

Massive stars dominate the spectral energy distributions (SEDs) of star-forming galaxies. 
Throughout their lives and during their explosive deaths as supernovae, they deposit energy, momentum, and ionizing photons into the surrounding interstellar medium (ISM).
This feedback from massive stellar populations is widely recognized as an important regulator of star formation and galaxy evolution processes \citep[e.g.,][]{somerville15, naab17}.
Bursts of star formation can efficiently create low-density channels in the ISM through which ionizing photons produced by young, massive stars can escape, particularly from low-mass galaxies in the early universe \citep[e.g.,][]{wise09, trebitsch17}.
Thus, massive stars in metal-poor dwarf galaxies are likely drivers of the reionization of the universe at high redshift, $z \gtrsim 6$ \citep{ouchi09, robertson15, finkelstein19}, but their ionizing fluxes remain uncertain.

The neutral intergalactic medium is opaque to H-ionizing photons ($\lambda \leq 912$\,\AA{}) during the epoch of reionization \citep[e.g.,][]{becker01}.
Observations of high-$z$ galaxies at longer rest wavelengths must therefore be modeled with stellar population synthesis (SPS) codes to infer the ionizing flux from massive stars.
SPS models combine template stellar spectra, stellar evolution models, an initial mass function (IMF), and a star formation history (SFH) parameterization to predict the intrinsic SED of a stellar population \citep[][]{tinsley80, conroy13}.
From this, the ionizing flux from massive stars can be inferred, then combined with observational constraints on ionizing photon escape fractions \citep[e.g.,][]{izotov16, steidel18} and the galaxy luminosity function at high $z$ \citep[e.g.,][]{livermore17, oesch18} to predict the total ionizing photons escaping from the predominantly metal-poor galaxies in the early universe.
Our understanding of the contribution of galaxies to cosmic reionization is therefore highly sensitive to the stellar evolution models and theoretical spectra adopted in SPS codes.

At solar metallicity (\zsun{}), a large population of massive stars in the Milky Way has been spectroscopically observed and forms a useful benchmark for calibrating stellar models \citep[e.g.,][]{howarth89, walborn90, przybilla10}.
The nearest star-forming dwarf galaxies that host sub-\zsun{} massive stars, the Large Magellanic Cloud (LMC; 50\%\,\zsun{}) and Small Magellanic Cloud (SMC; 20\%\,\zsun{}; \citealt{dufour84}), have been used to test the predictions of stellar models over a range of metallicity ($Z$; e.g., \citealt{mokiem07, dorn-wallenstein20}), but few spectroscopic observations of more metal-poor massive stars exist due to their large distances.
Even individual stars in ``nearby'' dwarf irregular galaxies just outside the Local Group are very faint and therefore expensive to observe, leaving us with few empirical constraints on stellar models appropriate to interpret observations of metal-poor galaxies in the early universe.

There is evidence from nearby metal-poor galaxies that stellar models are incomplete at low $Z$.
SPS models combined with high-quality SFHs inferred from resolved stellar populations of dwarf galaxies predict an excess of far-ultraviolet (FUV) flux over what is observed, even when the near-ultraviolet (NUV) flux is matched well \citep{mcquinn15a}.
In vigorously star-forming metal-poor galaxies, extreme nebular emission lines with strengths comparable to those observed in high-$z$ galaxies cannot be explained by the ionizing fluxes predicted by SPS models fit to the integrated light from the massive stellar populations \citep[e.g.,][]{berg18, berg19, senchyna19}.
A variety of possible solutions have been proposed, including binary interactions, a top-heavy IMF, and problems with theoretical spectra and/or evolution models at low $Z$ \citep[e.g.,][]{mcquinn15a, kehrig18, gotberg19, senchyna21}, but current observations cannot distinguish among these possibilities.
Empirical constraints on the SEDs and evolution of low-$Z$ massive stars are required to identify the cause of these model discrepancies and improve our ability to model metal-poor galaxies, both nearby and at high $z$.

\subsection{Constraints on the Astrophysics of Massive Stars from FUV Spectroscopy}

Predictions of stellar evolution models depend sensitively on the initial rotation speeds and prescriptions for mass-loss rates (\mdot{}) via stellar winds that they adopt.
Both affect the time evolution of massive stars' surface properties and their main-sequence lifetimes, which in turn set the total number of ionizing photons that a stellar population is predicted to produce \citep[e.g.,][]{levesque12, smith14}.
Theoretical grids of hot stellar spectra calculated with atmosphere modeling codes predict the metal opacities, wind line strengths, and ionizing fluxes of massive stars \citep[e.g.,][]{lanz03, eldridge17, martins21}. 
Below 20\%\,\zsun{}, these assumptions and predictions remain unconstrained by observations, so new measurements of the stellar and wind properties of individual OB stars are needed. 

FUV spectroscopy is a particularly important tool for measuring the properties of OB stars, as this spectral region contains many transitions of highly ionized metals and falls close to the peak of hot stellar SEDs.
The same metals, particularly Fe, that remove flux in the FUV also absorb strongly in the ionizing extreme UV part of the spectrum, so the observed opacities of metal lines in the FUV provide important constraints on the ionizing photon production by massive stars.
Metal lines absorb the emergent flux from hot stars to accelerate the expanding wind, forming P-Cygni profiles in the FUV that provide sensitive diagnostics of the terminal velocity (\vinf{}) and \mdot{} \citep{lamers99}.

At low $Z$, line-driven stellar winds are expected to weaken due to reduced metal opacities, though theoretical mass-loss prescriptions differ in the absolute \mdot{} for a given set of stellar parameters and predict somewhat different scalings with $Z$ \citep[e.g.,][]{vink01, bestenlehner20, bjorklund21, vink21}.
Observations in the Milky Way and Magellanic Clouds support the expectation of decreasing wind strength down to 20\%\,\zsun{} \citep[e.g.,][]{mokiem07}, though some empirical studies have found even weaker winds than expected in the SMC \citep{bouret13, ramachandran19}.
Measurements of \mdot{} and \vinf{} are rare for more metal-poor stars, so the question of whether these wind parameters obey the same scalings down to very low $Z$ remains open \citep{tramper11, tramper14, garcia14, bouret15}.

Winds can remove angular momentum from stellar surfaces throughout their lifetimes.
Since line-driven winds weaken at low $Z$, metal-poor stars are expected to lose angular momentum less efficiently and thus maintain higher rotation rates throughout their main-sequence lifetimes \citep[e.g.,][]{groh19}.
Observations of many OB stars in the Galaxy, LMC, and SMC hint that typical \vsini{} increases with decreasing metallicity \citep[e.g.,][]{penny09, ramachandran19}.
However, only a handful of measurements of \vsini{} or spectral line broadening have been reported below 20\%\,\zsun{} \citep{tramper11, tramper14, garcia14, ramachandran21}, so it remains unclear whether this trend holds for more metal-poor massive stars.
Fortunately, \vsini{} can be constrained from the structure of photospheric lines in moderate-resolution FUV spectroscopy \citep[e.g.,][]{penny09, bouret13}.

The ongoing \textit{Hubble Space Telescope} (\hst{}) Director's Discretionary Ultraviolet Legacy Library of Young Stars as Essential Standards (ULLYSES) program will measure new high-SNR, medium-to-high resolution (${R \sim 3000-45000}$) FUV spectra for over 160 OB stars in the LMC and SMC (spanning all luminosity classes for spectral types O2-B1.5, as well as B2-B9 supergiants).
These new data will be combined with archival observations to construct a library of FUV spectra for over 230 OB stars, providing an unprecedented benchmark for massive stellar astrophysics down to 20\,\%\,\zsun{}.
Yet, ULLYSES will only obtain low-resolution (${R \sim 1500-4000}$) FUV spectra of 6 OB stars in two more distant and metal-poor galaxies, NGC3109 (18\%\,\zsun{}; \citealt{lee07}) and Sextans~A (6\%\,\zsun{}; \citealt{skillman89}).
This will significantly add to the dozen published FUV spectra of OB stars in metal-poor dwarf galaxies outside of the Local Group as of this writing \citep{garcia14, garcia17, bouret15}, but this small sample is not sufficient to benchmark stellar models below 20\,\%\,\zsun{}.
Additional spectroscopic observations are needed to test SPS models, which are central to the interpretation of upcoming \textit{James Webb} and \textit{Nancy Grace Roman Space Telescope} observations of high-$z$ galaxies. 

Here, we present new FUV spectra of three metal-poor O~V stars in the nearby dwarf galaxies \object{Leo~P}, \object{Sextans~A}, and \object{WLM} ($3-14\%$\,\zsun{}), obtained with the Cosmic Origins Spectrograph (COS) on \hst{}. 
The paper is organized as follows: Section~\ref{sec:data} describes the sample of O stars and both spectral and photometric data used in this analysis. 
Section~\ref{sec:results} presents our analysis of the new FUV spectra and SEDs constructed with \hst{} photometry, including measurements of photospheric line strengths, wind line strength and velocity extent, projected rotation speed \vsini{}, and constraints on fundamental stellar properties from SED modeling.
Section~\ref{sec:discussion} discusses our results in comparison to analog stars in the SMC and implications for modeling metal-poor stellar populations.
Section~\ref{sec:conclusions} summarizes our conclusions. 

We define the terms that we use throughout the paper to refer to two distinct metallicity regimes. 
First, metal poor or low $Z$ indicates $Z < 20$\%\,\zsun{}, i.e., more metal-poor than the SMC. 
Second, we use the terms extremely metal poor and very low Z to mean $Z  < 10$\%\,\zsun{} (following the definition of very metal-deficient galaxies by \citealt{kunth00}).
We also consider the latter category to be substantially more metal-poor than the SMC (by at least a factor of two).


\section{Observations and Data Reduction\label{sec:data}}

\begin{table}
\tabcolsep=0.15cm
\begin{center}
\caption{Properties of the galaxies hosting the target stars.\label{tab:galaxies}} 
\begin{tabular}{llccc}
Star & Galaxy & 12+$\log$(O/H) & $Z/Z_\odot$ & Distance \\
 & & (nebular) & (nebular) & (Mpc) \\
\hline
LP26 & Leo P & $7.17\pm0.04$ & 0.03 & $1.62\pm0.15$  \\
S3 & Sextans A & $7.49\pm0.04$ & 0.06 & $1.38\pm0.05$ \\
A15 & WLM & $7.83\pm0.06$ & 0.14 & $0.98\pm0.04$ \\
\end{tabular}
\end{center}
\tablecomments{We adopt the star identifiers LP26, S3, and A15 from \citet{evans19}, \citet{garcia19}, and  \citet{bresolin06}, respectively. Nebular metallicities for Leo~P, Sextans~A, and WLM from \citet{skillman13}, \citet{skillman89}, and \citet{lee05}, respectively, are converted to solar units assuming solar abundance ratios and ($12+\log$(O/H))$_\odot = 8.69$ \citep{asplund09}. 
Distances to the galaxies (in the same order) are from \citet{mcquinn15b}, \citet{dalcanton09}, and \citet{jacobs09}. }
\end{table}

\begin{table*}
\begin{center}
\caption{Basic data for the three target stars and \hst{}/COS spectra. \label{tab:obs}}
\begin{tabular}{llcccccccc}
Star & Galaxy & RA & Dec. & $m_V$ & $A_V$  & Spectral & \multicolumn{2}{c}{Exposure Times (s)}  & Galaxy \vsys{} \\
 & & (J2000) & (J2000) & (mag) & (mag) & Type & G130M & G160M & (km\,s$^{-1}$) \\ 
\hline
LP26 & Leo P& 10:21:45.1217 & +18:05:16.93 & 21.51 & 0.04 & O7-8 V & 21502 & 48396 & 248 \\
S3 & Sextans A & 10:10:58.1866 & $-$04:43:18.45 & 20.80 & 0.09  & O9 V & 10725 & 26908 & 302 \\
A15 & WLM & 00:02:00.5333 & $-$15:29:52.41 & 20.25 &  0.04 & O7 V((f)) & 7880 & 15698 & $-115$ \\
\end{tabular}
\end{center}
\tablecomments{Right ascension is reported in hours, minutes, seconds, and declination is in degrees, arcminutes, arcseconds. 
Foreground extinction $A_V$  is from \citet{green15}. Visual magnitudes and spectral types for LP26, S3, and A15 are from \citet{mcquinn15b}, \citet{garcia19}, and \citet{bresolin06}, respectively.}
\end{table*}

In this section, we present the sample of O stars analyzed in this work. We then describe the observations and data reduction for both the new \hst{}/COS FUV spectra and the archival \hst{} imaging used to measure the broadband photometric properties of the stars. 

\subsection{The Sample of O~V Stars}

The goal of this work is to characterize empirically the FUV properties of metal-poor O stars still on the main sequence.
We therefore focus on known O stars with luminosity class V (i.e., dwarfs), as these are the least evolved.
Only one early-O dwarf in a galaxy more metal-poor than the SMC has ever been analyzed in the FUV, and that star (in IC~1613) appears to have an SMC-like Fe abundance \citep{bouret15}.
We searched the literature for all spectroscopically confirmed O~V stars residing in galaxies with gas-phase abundances below that of the SMC that are close enough that stars can be individually resolved and observed with \hst{}/COS \citep{bresolin06, evans19, garcia19}.
Excluding two stars in Sextans~A that will be observed as part of ULLYSES, our final sample consists of the only three known O~V stars that matched our metallicity and distance requirements.

Our target stars are LP26 in Leo~P (3\%\,\zsun{}), S3 in Sextans~A (6\%\,\zsun{}), and A15 in WLM (14\%\,\zsun{}).
LP26 is the only known O star in Leo~P, which is the closest star-forming galaxy with such low gas-phase metallicity. 
Host galaxy properties are given in Table~\ref{tab:galaxies}, and the spectral types and optical magnitudes of the stars are given in Table~\ref{tab:obs}.
We use the gas-phase oxygen abundances as estimates of the stellar $Z$ throughout this work, though it is likely that the true metal mass fractions of the stars are different as the $\alpha$/Fe ratio is known to vary across galaxies depending on the recent SFH.
Detailed stellar abundance measurements are outside the scope of this work and we defer that analysis to a future paper.
All three targets are considered metal-poor, and S3 and LP26 are extremely metal-poor.


\subsection{Photometry from Archival \hst{} Imaging\label{sec:imaging}}

\begin{figure*}[!htp]
\begin{centering}
\minipage{0.33\textwidth}
  \includegraphics[width=\linewidth]{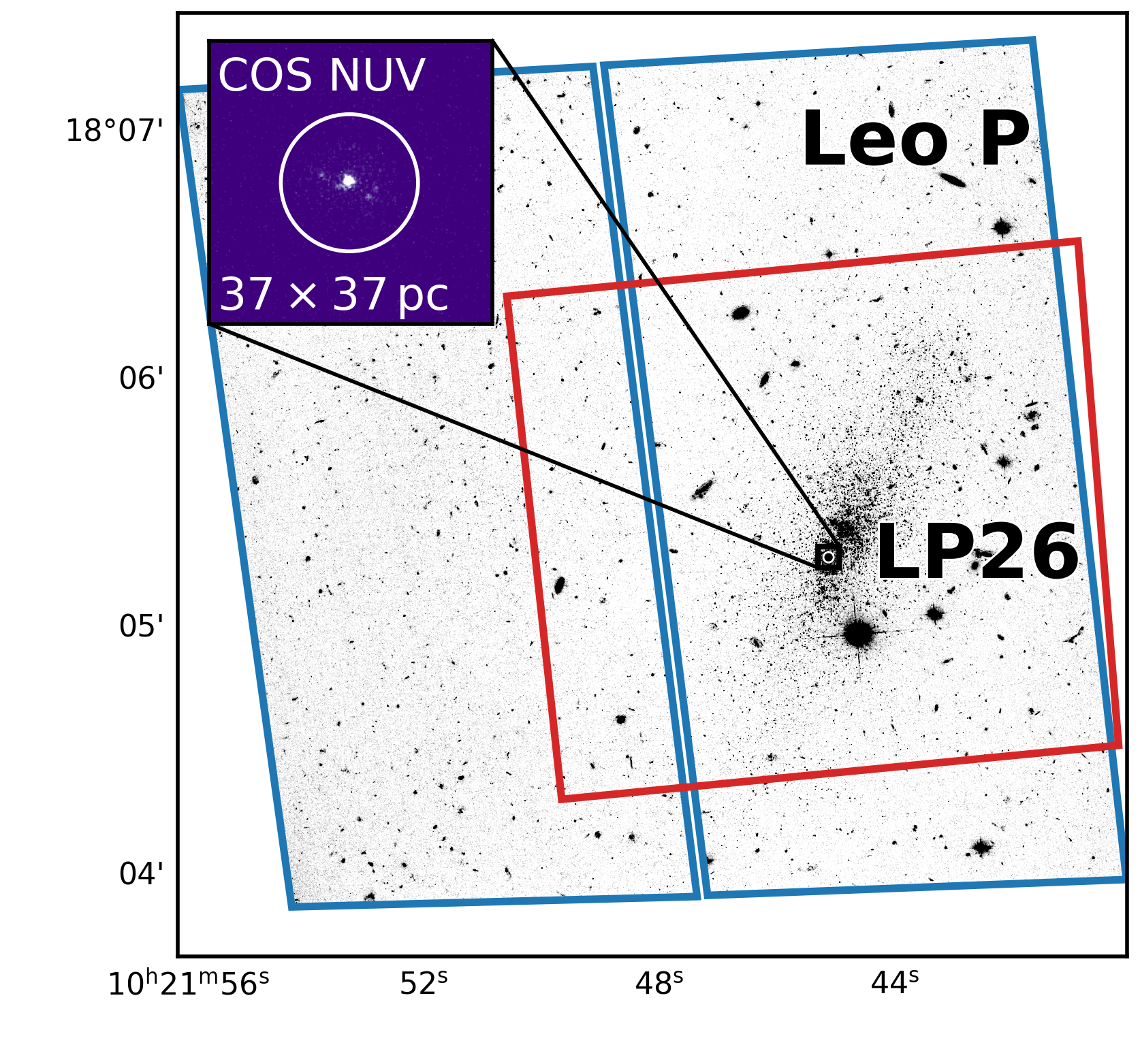}
\endminipage\hfill
\minipage{0.33\textwidth}
  \includegraphics[width=\linewidth]{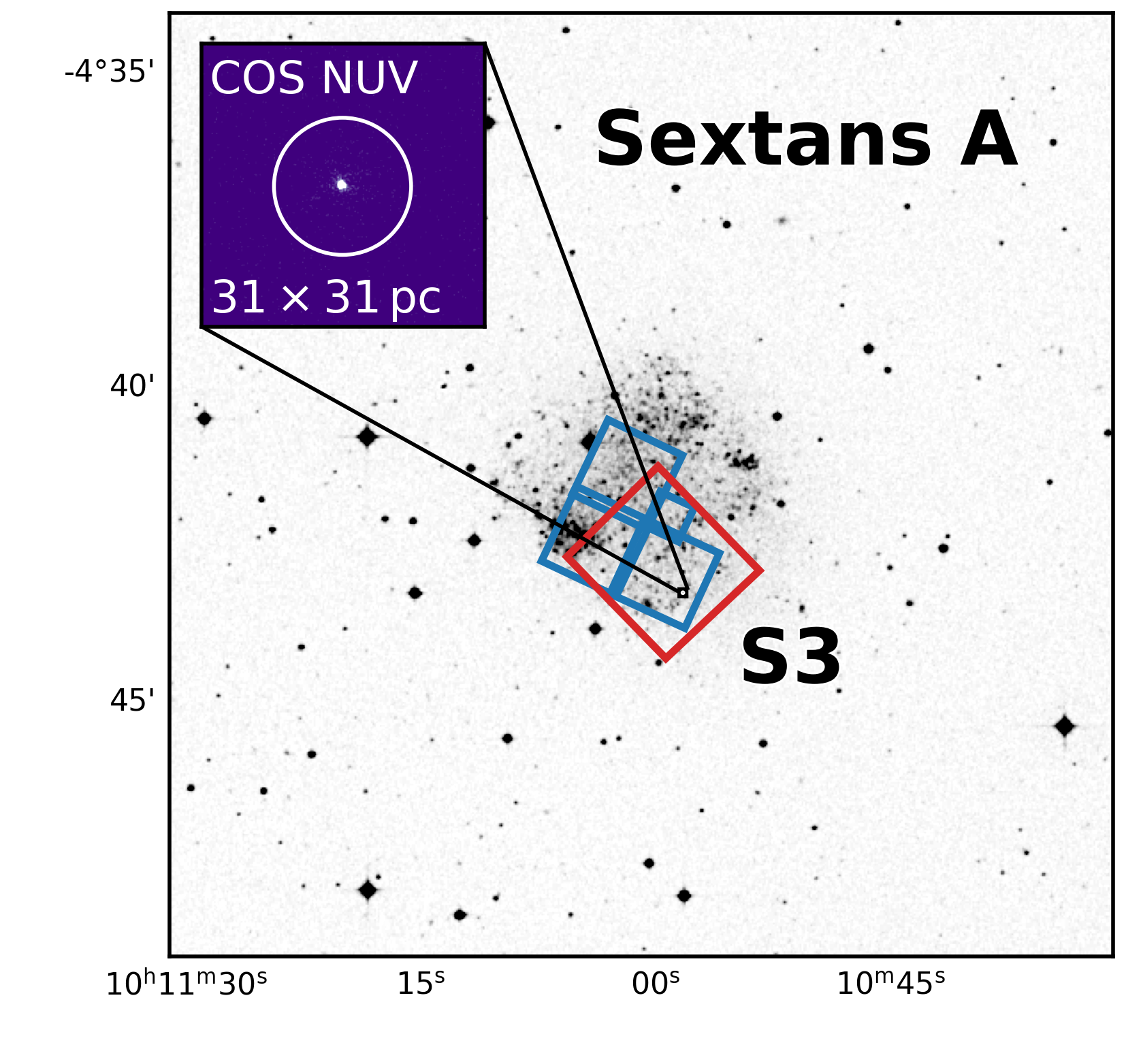}
\endminipage\hfill
\minipage{0.33\textwidth}
  \includegraphics[width=\linewidth]{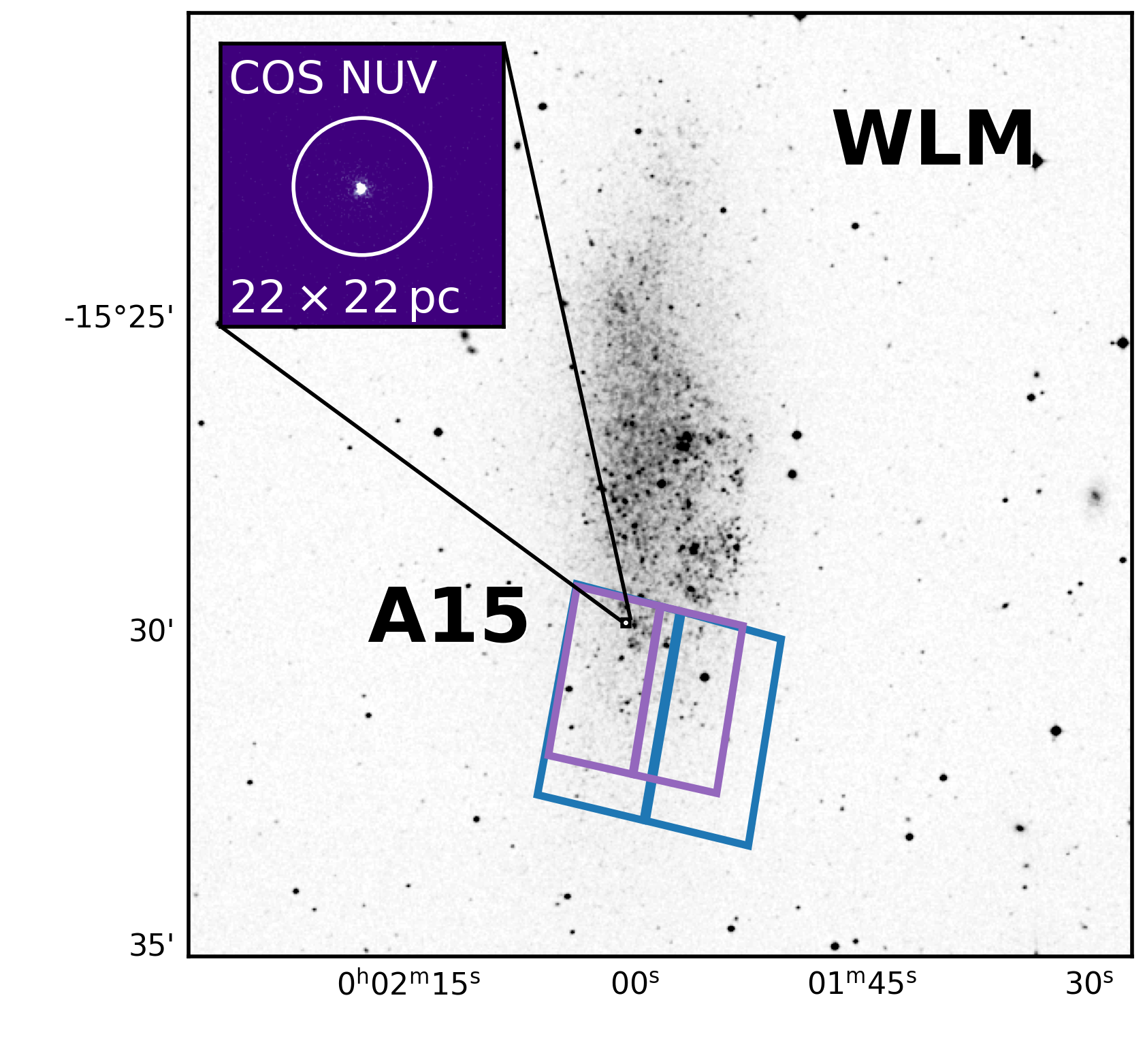}
\endminipage\hfill
\caption{\textbf{\textit{HST} imaging of the three target O stars in nearby, metal-poor galaxies.} From left to right, optical images of Leo~P, Sextans~A, and WLM are shown in greyscale with a linear stretch, all oriented with north up and east left. The Leo~P image is a $3\farcm{}4 \times 3\farcm{}4$  \hst{}/ACS F475W mosaic, while the other two are $15 \arcmin \times 15 \arcmin$ DSS images. The purple, blue, and red rectangles show the footprints of the archival \hst{} imaging in the NUV, optical, and NIR, respectively, that we use for stellar photometry. The insets centered on the three target stars show cutouts of the COS acquisition images $4\farcs{}7$ on a side, with the $2\farcs{}5$-diameter aperture shown as a white circle. Each cutout is annotated with its physical dimensions at the distance of the target star.
\label{fig:imaging}}
\end{centering}
\end{figure*}

\begin{table*}
\begin{center}
\caption{Photometry measured from archival \hst{} imaging. \label{tab:imaging}}
\tabcolsep=0.35cm
\begin{tabular}{lHHH|cc|ccc|ccc}
Star & Galaxy & Program ID & Filters & F275W & F336W & F475W & F555W & F814W & F127M & F139M & F153M \\
 &  &  &  & (mag) & (mag) & (mag) & (mag) & (mag) & (mag) & (mag) & (mag) \\ 
\hline
LP26 & Leo P& 13376 & F475W, F814W & \nodata & \nodata & 21.513 & \nodata & 21.929 & 22.312 & 22.466 & 22.341 \\
 & & 14845 & F127M, F139M, F153M & \nodata & \nodata & (0.002) & \nodata & (0.004) & (0.064) & (0.091) & (0.107) \\
 \hline
S3 & Sextans A & 5915 & F555W, F814W & \nodata & \nodata & \nodata & 20.866 & 21.082 & 21.492 & 21.417 & 21.400 \\
 &  & 14073 & F127M, F139M, F153M & \nodata & \nodata & \nodata & (0.005) & (0.009) & (0.043) & (0.036) & (0.044) \\
\hline
A15 & WLM & 15275 & F275W, F336W & 17.788 & 18.357 & 20.145 & \nodata & 20.526 & \nodata & \nodata & \nodata \\
 &  & 13768 & F475W, F814W & (0.003) & (0.002) & (0.001) & \nodata & (0.002) & \nodata & \nodata & \nodata \\
\end{tabular}
\end{center}
\tablecomments{Magnitudes are reported in the Vega system, and the uncertainty is given in parentheses below each measurement. These are the raw \dolphot{} outputs and are not corrected for dust extinction. Photometry was measured from \hst{} images from the following programs: 13376 (optical; PI: K. McQuinn) and 14845 (NIR; PI: M. Boyer) for LP26; 5915 (optical; PI: E. Skillman) and 14073 (NIR; PI: M. Boyer) for S3; and 15275 (NUV; PI: K. Gilbert) and 13768 (optical; PI: D. Weisz) for A15.}
\end{table*}

We require multi-wavelength photometry that samples the SEDs of the target stars as broadly in wavelength as possible to constrain their fundamental properties (Section~\ref{sec:sed_fitting}). 
All of our target stars are covered by high-resolution \hst{} imaging in the NUV, optical, and/or near-infrared (NIR), drawn from the programs listed in the caption of Table~\ref{tab:imaging}.
NUV and NIR images were taken with the Wide-Field Camera 3 (WFC3) UVIS and IR channels, respectively.
Optical imaging was taken with the Advanced Camera for Surveys (ACS) Wide Field Camera (WFC) detector for both Leo~P and WLM, and with the Wide-Field Planetary Camera 2 (WFPC2) for Sextans~A.
Both ACS datasets acquired many more images, which reach much fainter magnitudes when combined, than required to reliably measure the bright O star targets. 
We therefore only use 4 images per filter, taken during the same visit to ensure the best possible alignment.
Raw images (*flc.fits, *flt.fits, *c0m.fits, and *c1m.fits files, naming convention depending on the instrument) are retrieved from the Mikulski Archive for Space Telescopes (MAST). 

Figure~\ref{fig:imaging} shows optical images of each galaxy hosting one of our target stars in greyscale with the locations of the various \hst{} imaging datasets overlaid.
The $15\arcmin \times 15 \arcmin$ background images for WLM and Sextans~A are taken from the Digitized Sky Survey (DSS). 
Leo~P is smaller and fainter than ``classical'' dwarf galaxies and is not visible in DSS imaging, so for that galaxy we show an ACS F475W image, approximately $3\farcm4$ on a side.
The inset image in shades of purple in each panel shows an example COS acquisition image from our program, discussed further in Section~\ref{sec:cos_spectra}.
Outlines of the various archival \hst{} imaging footprints that we use to measure photometry for our target stars are overlaid, with purple, blue, and red outlines corresponding to NUV, optical, and NIR, respectively. 

We perform point spread function (PSF) photometry on the \hst{} images using \dolphot{} \citep{dolphin00, dolphin16}.
First, we use the tools in \textsc{drizzlepac} v3.1.8 \citep{gonzaga12} to clean cosmic rays from the images, align the individual images for each galaxy to a common reference frame, and create a deep mosaic in a filter in the middle of the wavelength range spanned by the data (F814W for Sextans~A and Leo~P, and F475W for WLM).
\dolphot{} uses the deep mosaic as a reference to which individual images are aligned and to locate stars.
The stars are then photometered in the individual images, and the measurements and uncertainties for each star are combined across all exposures in each filter.
Photometry is done simultaneously across all filters for each galaxy.
Table~\ref{tab:imaging} reports the measured photometry for each target star in the Vega magnitude system, in which the star Vega is defined to have zero magnitude at all wavelengths, and gives the uncertainties in parentheses below the measurements. 

The parameters set in \dolphot{} strongly affect the quality of the resultant photometry.
We generally follow the well-tested parameters used by the Panchromatic Hubble Andromeda Treasury (PHAT) survey \citep{williams14, williams21}. 
However, we find that using \textsc{FitSky}\,$=$\,2 (which is best for crowded stellar fields) results in our large, bright target stars being divided into multiple sources. 
Thus, we adopt \textsc{FitSky}\,$=$\,3 and the recommended values of \textsc{RAper} and \textsc{RPSF} for that setting from the instrument-specific \dolphot{} manuals.

In addition to measured flux and uncertainty for each star, \dolphot{} output includes the quality metrics \textsc{crowd} and \textsc{sharp}. 
High \textsc{crowd} values indicate that the star's photometry is potentially affected by the presence of nearby sources, while high $|$\textsc{sharp}$|$ indicates that the source is more extended than expected for a star (e.g., background galaxy) or too narrow (e.g., cosmic ray).
We verify that high-quality photometry is obtained for all three target stars; specifically, \textsc{crowd}\,$\leq$\,0.45\,mag and $|$\textsc{sharp}$|^2$\,$\leq$\,0.05 in all filters.
Typical photometric signal-to-noise ratios (SNRs) reach several hundred in the NUV and optical and $\sim 10-30$ in the NIR. 

We correct the photometric fluxes for Milky Way foreground dust extinction adopting $A_V$ at the location of each star from \citet{green15} and a \citet{fitzpatrick99} extinction law with $R_V = 3.1$.
The shape of the stellar SED affects the magnitude of the extinction correction in each filter. 
We calculate the corrections for the SED of a 40\,kK dwarf to ensure suitability for the O stars in this study.


\subsection{\hst{}/COS FUV Spectra\label{sec:cos_spectra}}

We obtained medium-resolution FUV spectra of our targets with \hst{}/COS in Cycle 27 over 49 orbits between 2020 April 10--June 16. (GO-15967; PI: J. Chisholm).
Table~\ref{tab:obs} gives basic data for the three target stars and exposure times of the observations.  
We used the G130M grating at central wavelength 1291\,\AA{} (FP-POS=3, 4) and the G160M grating at central wavelength 1600\,\AA{} (FP-POS=all). 
This combination covers many lines that serve as diagnostics of stellar photosphere and wind properties, particularly the \trans{C}{iii}{1176} and \trans{N}{v}{1240} lines in G130M and \trans{C}{iv}{1550} in G160M. 
Observations were taken in TIME-TAG mode using the $2\farcs{}5$ Primary Science Aperture (PSA). 
The calibrated and co-added spectra for all 12 visits (*x1dsum.fits files), processed and extracted with the standard CalCOS pipeline (version 3.3.9), were downloaded from MAST. 
We verified that the wavelength solutions and flux calibrations using the default parameters were consistent across visits. 

Example COS acquisition images for each target star, taken in the NUV channel, are shown in the inset panels of Figure~\ref{fig:imaging}.
Each inset is a $4.7 \arcsec$ on a side cutout of the acquisition image, and we annotate the insets with the corresponding physical size at the distance of each galaxy (ranging from $22-37\,\mathrm{pc}$ on a side). 
We show the $2\farcs{}5$-diameter COS PSA as a white circle, verifying that all three stars are well-centered and isolated in the aperture, so the extracted COS/FUV spectra will accurately measure photons from the target stars.
All target stars are approximately point sources as expected, though some diffuse emission surrounding LP26 is visible in the NUV acquisition image, likely due to the \hii{} region in which the star is embedded.

The three target stars are distant and therefore faint, so multiple visits (up to 5 per grating) were typically required to achieve high enough SNR to detect weak photospheric absorption lines in the continuum. 
For each star, we resample all spectra onto a common wavelength grid and coadd the spectra across all visits, weighting by the inverse variance. 
This procedure both increases the SNR and produces a single composite spectrum for each star. 
The dispersion of the combined spectra is 12.23\,m\AA{}/pixel, limited by the dispersion of the G160M grating\footnote{\url{https://hst-docs.stsci.edu/cosihb}}. 

We measure the spectral resolution as the FWHM of Gaussian line profiles fit to the Milky Way ISM absorption features in the FUV spectra, specifically \trans{si}{ii}{1190,\,193,\,1260, 1526} and \trans{c}{ii}{1134}. 
Across all three stars we find typical velocity FWHM of $\sim50\,\mathrm{km\,s}^{-1}$, corresponding to $R\sim6000$. 
This measured FWHM is about three times larger than the theoretical 6-pixel resolution element of $\sim15\,\mathrm{km\,s}^{-1}$. 
Individual absorption components due to the Milky Way and target galaxy ISM can be resolved in the spectra, enabling us to measure the systemic velocity \vsys{} of each host galaxy (reported in Table~\ref{tab:obs}).
We choose to bin the coadded spectra by 12 pixels ($\sim30\,\mathrm{km\,s}^{-1}$) for all analysis to boost the SNR as much as possible while ensuring $1.5-2$ samples per resolution element across the entire spectral range. 
This binning achieves SNR between $6.5-10.3$, $9.6-12.7$, and $5.6-7.3$ at 1170\,\AA{}, 1430\,\AA{}, and 1560\,\AA{}, respectively, with the S3 data reaching the lowest SNR and A15 the highest across the entire FUV wavelength range.
Finally, we correct the spectra for Milky Way foreground dust extinction assuming a \citet{fitzpatrick99} extinction law with $R_V = 3.1$.


\section{Analysis of FUV Spectra and Spectral Energy Distributions\label{sec:results}}

Here, we present and analyze the new FUV spectra of three low-$Z$ O stars. These include the first such observations of O dwarfs below 10\%\,\zsun{} (LP26 and S3), enabling new constraints the properties of extremely metal-poor, unevolved massive stars. We quantify the strengths of photospheric and wind lines, estimate their projected rotation speeds \vsini{}, and constrain the stars' fundamental parameters via SED fitting.


\subsection{FUV Metal Opacities\label{sec:opacities}}

\begin{figure*}[!ht]
  \includegraphics[width=\linewidth]{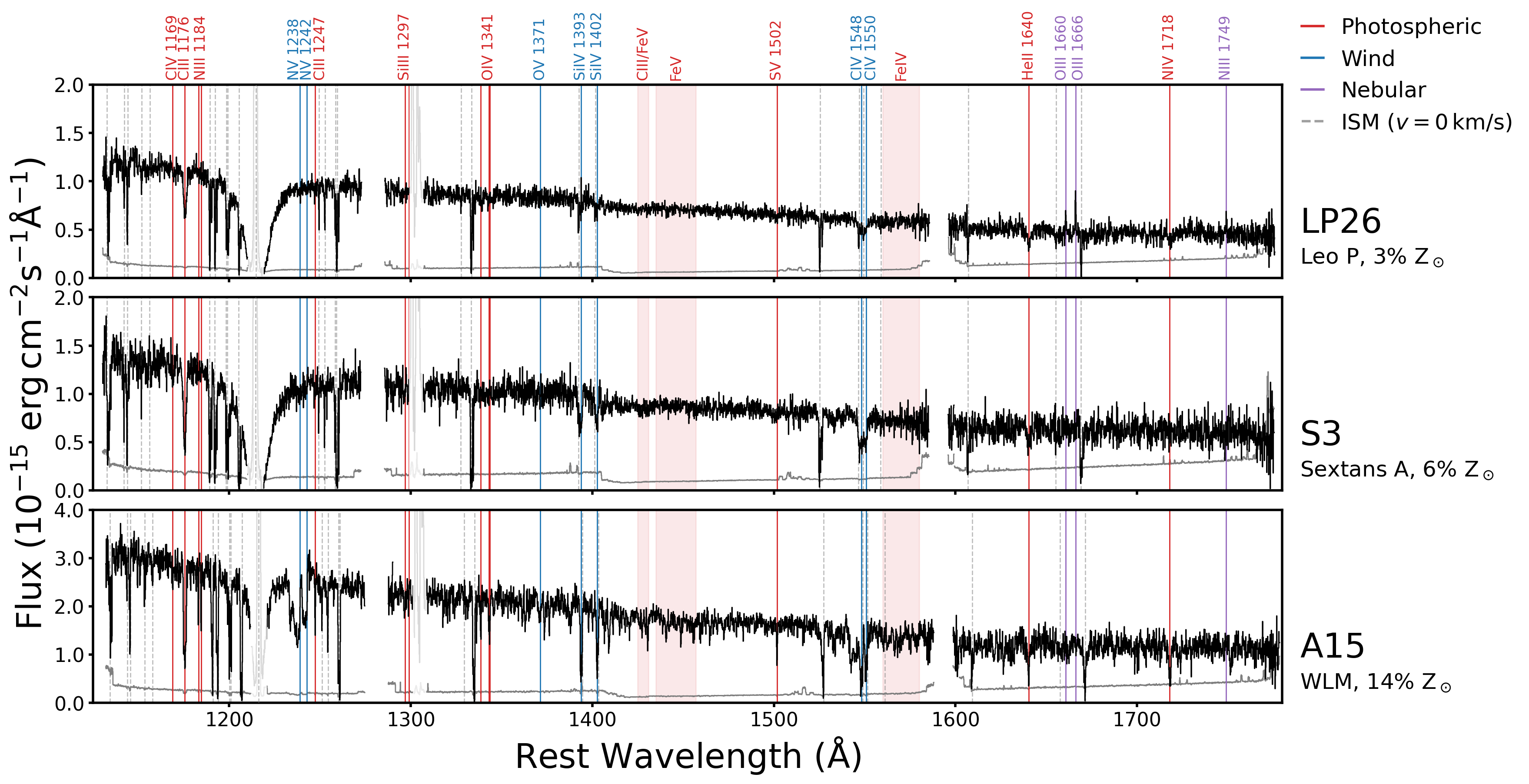}
\caption{\textbf{\textsc{The first FUV spectra of extremely metal-poor O-dwarf stars.}} The \hst{}/COS spectra of our target stars in Leo~P (top), Sextans~A (middle), and WLM (bottom) are plotted in black as a function of wavelength, corrected to the host galaxy rest frame using the systemic velocity measured from the ISM lines. The uncertainties are shown as the solid grey line in each panel. Spectral regions affected by geocoronal emission lines have been greyed out for clarity, and the locations of Milky Way ISM absorption lines are shown by the grey dashed lines. Stellar photospheric and wind features are indicated by red and blue vertical lines, respectively, and the nebular emission lines visible in the LP26 spectrum are shown by the purple lines. Red shading highlights ``forests'' of primarily Fe transitions. The increasing line blanketing by Fe-peak elements and increasing wind strength from 3\% to 14\% \zsun{} is evident by eye. The LP26 star is the lowest-$Z$ star ever observed in the FUV, and these data reveal a surprisingly featureless continuum as well as rare nebular emission lines powered by a single O star.
\label{fig:spectra}}
\end{figure*} 

\begin{table*}
\caption{Equivalent Widths of FUV Photospheric Lines}
\label{tab:ews}
\begin{center}
\tabcolsep=0.35cm
\begin{tabular}{l|cc|cc|cc|c}

 \multicolumn{1}{c}{ } & \multicolumn{2}{c}{LP26} & \multicolumn{2}{c}{S3} & \multicolumn{2}{c}{A15} & All Stars \\
Feature & EW & $\sigma_\mathrm{EW}$ & EW & $\sigma_\mathrm{EW}$ & EW & $\sigma_\mathrm{EW}$ & Feature Bandpass \\
 & ($\mathrm{\AA}$) & ($\mathrm{\AA}$) & ($\mathrm{\AA}$) & ($\mathrm{\AA}$) & ($\mathrm{\AA}$) & ($\mathrm{\AA}$) & ($\mathrm{\AA}$) \\
\hline 
C\,\textsc{iii}\,1176* & 0.83 & 0.09& 1.26 & 0.13& 1.04 & 0.08& $1173.5-1177.5$ \\
N\,\textsc{iii}\,1184* & \nodata & 0.10 & \nodata & 0.14 & 0.24 & 0.08& $1182.5-1185.2$ \\
C\,\textsc{iii}\,1247 & 0.11 & 0.05& \nodata & 0.07 & 0.17 & 0.04& $1246.4-1248.0$ \\
O\,\textsc{iv}\,1341* & 0.34 & 0.17& \nodata & 0.23 & 0.36 & 0.16& $1337.5-1344.5$ \\
S\,\textsc{v}\,1502 & \nodata & 0.07 & \nodata & 0.10 & 0.21 & 0.07& $1500.5-1503.0$ \\
He\,\textsc{ii}\,1640 & 0.77 & 0.30& \nodata & 0.34 & 0.91 & 0.23& $1638.6-1643.0$ \\
N\,\textsc{iv}\,1718 & \nodata & 0.28 & \nodata & 0.34 & 0.75 & 0.23& $1716.6-1719.7$ \\
Fe\,\textsc{v}/C\,\textsc{iii}\,1428* & \nodata & 0.12 & \nodata & 0.15 & 0.73 & 0.10& $1425.0-1431.0$ \\
Fe\,\textsc{v}\,1445* & \nodata & 0.46 & \nodata & 0.63 & 1.46 & 0.41& $1435.0-1457.0$ \\
Fe\,\textsc{iv}\,1570* & \nodata & 0.96 & \nodata & 1.26 & 1.70 & 0.54& $1560.0-1580.0$ \\
\end{tabular}
\end{center}
\tablecomments{All features included in this table were detected at the 2$\sigma$ level in at least 1 star. No EW is reported for measurements with lower SNR; in those cases, $\sigma_\mathrm{EW}$ is taken to be an upper limit on the EW. Features marked with * contain multiple transitions of the same ion.}
\end{table*}

Many transitions of metal ions common in hot stellar atmospheres lie in the FUV, including forests of weak Fe lines that dominate the metal opacity.
The strengths of these lines are sensitive to the temperature, ionization structure, and metal abundances in the stellar atmospheres. 
Medium-resolution COS spectroscopy enables detection of even relatively weak lines expected at low $Z$ and empirically constrains the opacity sources in the target stars.
Figure~\ref{fig:spectra} presents the co-added and flux-calibrated FUV spectra of the three stars, shown as black lines. 
Wavelengths have been corrected to the rest frame using the systemic velocity measured for each galaxy from the Gaussian modeling of ISM absorption lines described in Section~\ref{sec:cos_spectra} (see Table~\ref{tab:velocities} below). 
The measurement uncertainties on the spectra are shown as the solid grey lines in each panel.
The colored vertical lines indicate the locations of various features in the spectra: photospheric absorption (red), stellar wind absorption and emission (blue), and nebular emission (purple). 
Grey dashed vertical lines show the locations of ISM absorption at the velocity of the Milky Way. 
Regions of the spectra contaminated by geocoronal emission lines have been greyed out.

The three FUV spectra are ordered from top to bottom by increasing nebular abundance of the host galaxy. 
By eye, it is obvious that the continuua of the lower-$Z$ LP26 and S3 are quite featureless compared to that of A15, though the most prominent photospheric lines are visible in all three spectra (e.g., \trans{c}{iii}{1176} and \trans{he}{ii}{1640}). 
The wind-sensitive lines are clearly much more prominent in A15 than in the two lower-$Z$ stars; we discuss this in detail in Section~\ref{sec:windlines} below.
Nebular emission lines are only present in the spectrum of the lowest-$Z$ star, LP26, as expected since this is the only target star embedded in an \hii{} region. 

We quantify the strengths of the photospheric lines by measuring their equivalent widths (EWs), defined as:
\begin{equation}
\mathrm{EW} = \int_\mathrm{bandpass} 1 - \left(\frac{F_\lambda}{F_\lambda^\mathrm{continuum}}\right).
\end{equation}
The feature bandpasses are chosen to cover the entire observed line profile across all three spectra and minimize contamination from nearby ISM and strong photospheric lines predicted by \tlusty{} theoretical spectra drawn from the \textsc{ostar2002} grid \citep{lanz03}. 
Table~\ref{tab:ews} reports the rest wavelengths of the adopted feature bandpasses, where the spectra are corrected for the velocity of the stars determined from cross-correlating observed photospheric lines with the transitions in the \tlusty{} model spectra (discussed further in Section~\ref{sec:velocities} below).
The continuum flux level is determined from $1-5$\,\AA{} regions on either side of the feature bandpass, similarly chosen to avoid contaminating ISM lines and stellar lines.
The continuum bandpasses adopted for all photospheric and wind features are reported in Table~\ref{tab:contnorm} in Appendix~\ref{sec:appendix}.
We perform a least-squares linear fit across the continuum bandpasses to estimate the local continuum level around each photospheric line. 
We resample the flux 1,000 times, adding offsets to the measured flux drawn from a Gaussian distribution with a mean of 0 and a standard deviation equal to the measured flux uncertainty. 
Continuum levels and EWs are re-measured for all resampled spectra, and the EW uncertainty ($\sigma_\mathrm{EW}$) is defined as the standard deviation of these EW measurements.

Table~\ref{tab:ews} presents the EW measurements for various photospheric lines in the three stars LP26, S3, and A15. 
Only EWs measured at the $2\sigma$ level are reported; if the SNR is lower, we consider that feature to be undetected and the $\sigma_\mathrm{EW}$ is taken to be an upper limit on its EW. 
Some bandpasses contain multiple transitions of the same ion, indicated with asterisks in Table~\ref{tab:ews}.
The last three rows of Table~\ref{tab:ews} report the EWs of wider ``forest" regions that contain a large number of transitions that can eat away at the continuum level: \trans{fe}{v}{1445}, Fe\,\textsc{v}/C\,\textsc{iii}\,1428, and \trans{fe}{iv}{1570} (red shaded regions in Figure~\ref{fig:spectra}). 
Again, we use \tlusty{} model spectra to select regions for continuum normalization expected to be free of photospheric line absorption (Table~\ref{tab:contnorm}).
The high density of lines in these forest regions restricts us to narrow ($\lesssim 2$\,\AA{}) continuum bandpasses, and these are not guaranteed to sample the true continuum level, though metal line absorption reducing the apparent continuum level should be less problematic at low $Z$.
Interestingly, we only detect the depletion of continuum in these Fe forests for the highest-metallicity star, A15.
This is a quantitative confirmation of the visual impression from Figure~\ref{fig:spectra} that the two very low-$Z$ stars have featureless continuua, even in the higher SNR $1400-1500$\,\AA{} range. 

Some lines that might be expected in these mid-late O stars \citep{heap06} are not detected in any of the three at the $2\sigma$ level, including \trans{c}{iv}{1169} and the \spec{si}{iii} triplet near 1296\,\AA{}.
This is due to a combination of the weakness of those lines in the low-$Z$ stellar spectra, the SNR of the spectra, and the difficulty of estimating a reliable continuum level for \trans{si}{iii}{1296} in particular due to its location between a gap in the spectra and prominent \spec{o}{i} geocoronal emission.
However, the \trans{c}{iv}{1169} line is in a relatively high SNR region, and upper limits on its EW range from 0.08 to 0.13\,\AA{}, with the smallest upper limit found for the highest-$Z$ star, A15.
The ratio of \trans{c}{iii}{1176} to \trans{c}{iv}{1169} is sensitive to \teff{} \citep{heap06}, so this non-detection of \trans{c}{iv}{1169} may suggest that all three target stars are relatively cool. 
We return to this point in Section~\ref{sec:sed_results} below.


\subsection{Stellar Winds\label{sec:windlines}}

\begin{figure*}[!ht]
  \includegraphics[width=\linewidth]{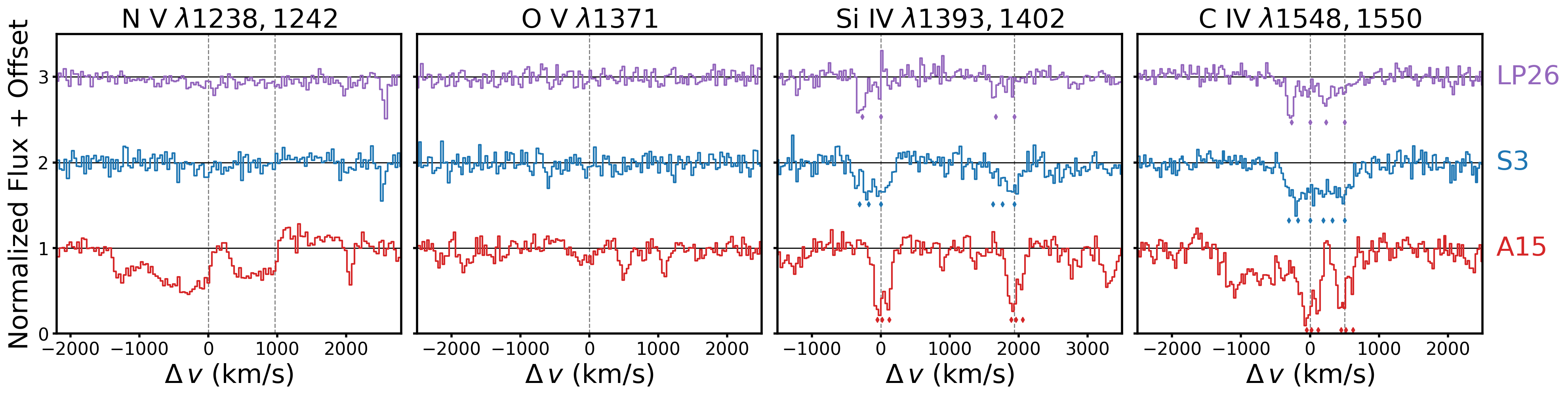}
\caption{\textbf{\textsc{Weak stellar wind features at extremely low metallicity.}} Continuum-normalized spectra are plotted as a function of velocity offset from the center of wind-sensitive lines, from left to right: \trans{n}{v}{1238},  \trans{o}{v}{1371}, \trans{si}{iv}{1393}, and \trans{c}{iv}{1548}. Vertical dashed lines indicate zero velocity for the bluer transition and the velocity of the redder transition in the doublets. The spectra of LP26 (purple), S3 (blue), and A15 (red) are presented in order of increasing $Z$ from top to bottom. Spectra are normalized to a continuum level of 1 (indicated by horizontal black lines), then vertically offset from each other for clarity. Diamonds below each spectrum indicate the velocities of ISM \spec{si}{ii} and \spec{c}{ii} absorption components detected in the spectrum of that star. Broad, blueshifted \spec{c}{iv} wind absorption is clearly present in S3 and A15, and possibly also in LP26. A15 also shows wind absorption and emission in the \spec{n}{v} and \spec{o}{v} lines. All stars' \spec{si}{iv} lines are likely due to photospheric and ISM absorption rather than winds.
\label{fig:windlines}}
\end{figure*} 

\begin{table*}
\caption{FUV Wind Line Properties}
\label{tab:windlines}
\begin{center}
\tabcolsep=0.35cm
\begin{tabular}{llccccc}
Star & Feature & $v_\mathrm{edge}$ & $\sigma_{v_\mathrm{edge}}$ & EW & $\sigma_\mathrm{EW}$ & Feature Bandpass \\
 & & (km\,s$^{-1}$) & (km\,s$^{-1}$) & ($\mathrm{\AA}$) & ($\mathrm{\AA}$) & ($\mathrm{\AA}$) \\
\hline 
S3 & C\,\textsc{iv}\,1548 & 440 & 120& \nodata & \nodata & \nodata \\
A15 & C\,\textsc{iv}\,1548 & 1370 & 150& \nodata & \nodata & \nodata \\
A15 & N\,\textsc{v}\,1238 & 1410 & 150& 2.12 & 0.10& $1233.0-1239.8$ \\
A15 & N\,\textsc{v}\,1242 & \nodata & \nodata & 0.83 & 0.06& $1239.9-1243.2$ \\
A15 & N\,\textsc{v}\,1242 Emission & \nodata & \nodata & $\leq -0.43$ & 0.07& $1243.2-1246.7$ \\
\end{tabular}
\end{center}
\tablecomments{EWs are defined such that absorption is positive and emission is negative. The A15 N\,\textsc{v}\,1242 emission EW is an upper limit beacuse the feature bandpass was truncated to avoid the blended C\,\textsc{iii}\,1247 photospheric line.}
\end{table*}

Resonance lines in the FUV are excellent stellar wind diagnostics, sensitive to both \mdot{} and velocity structure.
The shapes of the wind profiles give empirical insight into the strength and \vinf{} of a stellar wind.
P-Cygni wind profiles are produced by wind emission that extends to $\pm v\,_\infty$ and blueshifted absorption from $-v_\infty$ to zero velocity. 
\vinf{} can therefore be measured empirically from the blueward extent of the ``black part" of the absorption trough for fully saturated lines (i.e., absorption that reaches a flux of 0). 
For unsaturated P-Cygni profiles, the wind may not be optically thick to absorption in the outer, higher velocity regions, and so the blueward extent of the wind absorption represents a lower limit on \vinf{} \citep{lamers95, crowther16}.
The depth of the wind absorption troughs and presence of emission components indicate the density of the wind material, such that stronger wind profiles signal higher \mdot{}.
Quantitative measurement of \mdot{} requires computationally intensive atmosphere modeling that accounts for hydrodynamics, metal abundances, and ionization structure and is beyond the scope of the present work.
Here, we discuss the empirical constraints that FUV wind line profiles provide on winds that metal-poor O~V stars are capable of driving.

Figure~\ref{fig:windlines} shows the wind-sensitive lines in the COS spectra of the three target stars. Sections of the spectra of LP26 (top/purple), S3 (middle/blue), and A15 (bottom/red) are plotted against velocity relative to the wavelength of the bluest transition in each panel. 
The spectra have been normalized to a continuum level of 1, then plotted with an integer offset for clarity. 
We select continuum regions free of photospheric, wind, and ISM lines on either side of each wind-sensitive feature (reported in Table~\ref{tab:contnorm} in Appendix~\ref{sec:appendix}) and perform a linear fit to the local continuum for all lines except \spec{n}{v}, which requires a third-order polynomial to capture the shape of the red wing of Lyman\,$\alpha$ absorption. 
Vertical dashed lines indicate the relative velocities of each wind line transition, where $v=0$ is the velocity of the bluest transition in each panel, and diamonds below each spectrum show the velocities of narrow ISM components determined from modeling \spec{si}{ii} and \spec{c}{ii} lines (Section~\ref{sec:cos_spectra}). 

It is clear that A15, which is both the highest-$Z$ and earliest type of the three stars, has the strongest wind features. 
Broad, blueshifted absorption is present in both the \trans{n}{v}{1240} and \trans{c}{iv}{1549} lines, as well as redshifted emission in \spec{n}{v}. 
The \trans{o}{v}{1371} line also appears affected by the stellar wind, with a broad, blueshifted absorption feature bracketed by emission. 
This weaker, non-resonance transition may only absorb in the denser, lower-velocity part of the wind, leading to blueshifted emission not seen in resonance lines. 
The EW of \trans{o}{v}{1371} wind absorption has been used as a \teff{} diagnostic above $\sim40\,\mathrm{kK}$ \citep{dekoter98}, but the weak absorption strength in the A15 spectrum suggests a cooler \teff{} than that threshold.

The spectra of the two extremely metal-poor stars reveal weaker wind features.
S3 shows broad, blueshifted absorption in the \spec{c}{iv} line, but no evidence of wind features in any of the other transitions. 
LP26, the lowest-metallicity star, shows a hint of broad absorption in \spec{c}{iv}, but this feature is weak and its blue edge is contaminated by narrow ISM absorption. 
Overall, the lowest-$Z$ star in this sample does not show evidence of strong stellar winds, suggesting very low \mdot{} as expected for extremely metal-poor, main-sequence O stars.
None of the three spectra show any wind-like features in the \spec{si}{iv} lines, suggesting that their winds are hot and/or low-density \citep[e.g.,][]{chisholm19}. 

We quantify the strength and velocity extent of the clearly detected wind features in Table~\ref{tab:windlines}. 
LP26 is excluded from this analysis due to the very weak (if present at all) wind features in its FUV spectrum and contamination of the \spec{c}{iv} feature by strong ISM absorption.
We measure \vedge{} for A15 and S3, defined as the velocity where the blue edge of the wind absorption meets the normalized continuum level of 1. 
This is an approximation of \vinf{} for optically thick winds \citep[e.g.,][]{crowther16}, or for weaker winds, a measure of the velocity extent of the denser, inner part of the wind.
We measure \vedge{} of 450\,km\,s$^{-1}$ from the \trans{c}{iv}{1548} line in S3's spectrum, which is much lower than the expected \vinf{} for an O9~V star, even after accounting for the expected metallicity dependence $v_\infty \propto Z^{0.13}$ \citep{leitherer92}: scaling the typical \vinf{} for Galactic O9~V stars given by \citet{kudritzki00} gives an expected $v_\infty \sim 1040\,\mathrm{km\,s}^{-1}$ at 6\%\,\zsun{}.
This wind profile is likely under-developed and not optically thick in the outer, fastest regions of the wind, so the empirical \vedge{} measurement underestimates \vinf{} of the wind in S3.
A15 has two well-developed wind profiles at \trans{n}{v}{1238} and \trans{c}{iv}{1548}.
Averaging the measurements for the two lines, we find \vedge{} of 1390\,km\,s$^{-1}$.
The wind lines are not saturated, so this should be treated as an upper limit on \vinf{}.
Indeed, our measured \vedge{} is lower than the expected $v_\infty \sim 1860\,\mathrm{km\,s}^{-1}$ for an O7~V star at 14\%\,\zsun{}, again assuming the \citet{leitherer92} scaling.
On the other hand, it is consistent with reported \vinf{} measurements as low as $1300\,\mathrm{km\,s}^{-1}$ for O6.5-7~V stars in the SMC \citep[][]{bouret13}.

Table~\ref{tab:windlines} also reports the EWs of the \trans{n}{v}{1238,\,1242} absorption and \trans{n}{v}{1242} emission, which are only detected in the spectrum of A15. 
These lines are prominent and not contaminated by ISM absorption, though the absorption trough of the redder line is likely impacted by emission due to the bluer line. 
By convention, the EW of emission is negative, with lower values indicating stronger emission. The \trans{n}{v}{1242} emission is blended with the \trans{c}{iii}{1247} photospheric line, so we restrict our EW measurement to wavelengths blueward of that line, resulting in an upper limit on the EW (i.e., the emission is at least as strong as the reported EW indicates). 
Uncertainties on both EW and \vedge{} are estimated from repeating the measurements 1,000 times on spectra resampled from their uncertainty distributions, as explained in Section~\ref{sec:opacities}.


\subsection{Projected Rotation Velocities\label{sec:velocities}}

\begin{figure*}[!t]
  \includegraphics[width=\linewidth]{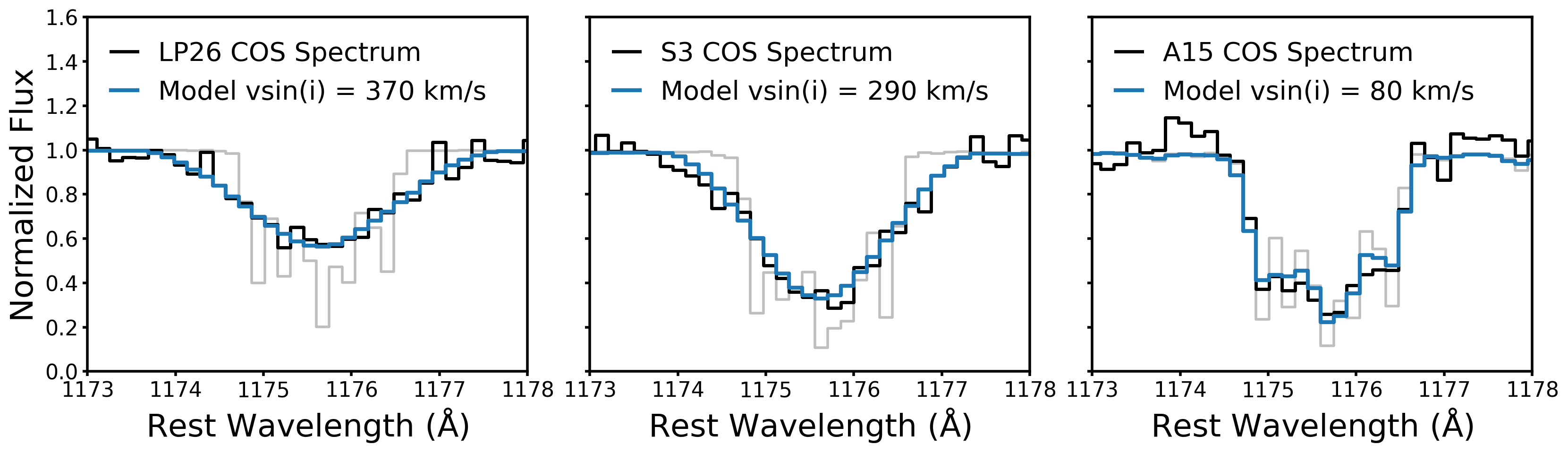}
\caption{\textbf{\textsc{Both extremely metal-poor stars are fast rotators.}} The continuum-normalized COS spectra around the complex of photospheric \spec{c}{iii} around 1176\,\AA{} are shown as black lines for LP26, S3, and A15 (left to right). The grey line in each panel shows the \tlusty{} model that most closely matches the measured \trans{c}{iii}{1176} equivalent width for each star with no rotation, broadened to COS resolution and resampled to the same wavelength grid as the data. The blue line in each panel shows the rotationally broadened \tlusty{} model that best reproduces the observed \spec{c}{iii} profile, and the best-fit \vsini{} for each star is reported in the legends. High \vsini{} is required to match the smooth and broad profiles in the LP26 and S3 spectra.
\label{fig:vsini}}
\end{figure*} 

\begin{table}
\tabcolsep=0.25cm
\begin{center}
\caption{Projected rotation velocities, radial velocities, and host galaxy systemic velocities measured from FUV spectra. Galaxy \vsys{} measurements are repeated from Table~\ref{tab:obs} to facilitate comparison with stellar \vrad{}. \label{tab:velocities}}
\begin{tabular}{lccc}
Star & \vsini{} & $v_\mathrm{radial}$ & Galaxy $v_\mathrm{sys}$ \\
 & (km\,s$^{-1}$) & (km\,s$^{-1}$) & (km\,s$^{-1}$) \\
\hline
LP26 & $370 \pm 90$ & 283 & 248 \\
S3 & $290 \pm 90$ & 296 & 302 \\
A15 & $80 \pm 45$ & $-145$ & $-115$ \\
\end{tabular}
\end{center}
\end{table}

The 50\,km\,s$^{-1}$ resolution of the COS spectra enables us to constrain the projected rotation speeds, \vsini{}, of our target stars.
Rotation broadens intrinsically narrow absorption lines, changing the shapes of the observed profiles but keeping the EW intact. 
\vsini{} can therefore be constrained by convolving appropriate theoretical spectra with a rotational broadening kernel to reproduce the observed photospheric line profiles \citep[e.g.,][]{huang06, bouret13}.
The \trans{c}{iii}{1176} complex is a particularly useful diagnostic, as the six closely spaced lines remain separable at \vsini{}\,$\lesssim 40$\,km\,s$^{-1}$ \citep{heap06}, and at higher \vsini{} the shape of the smoothed profile (particularly the wings) remains sensitive to changes in rotation.
This line is also the only photospheric feature detected at the 2$\sigma$ level in all three COS spectra, enabling a uniform measurement technique across the sample.
While \trans{c}{iii}{1176} can manifest as a wind line at high $Z$, previous work on SMC O stars has shown that this transition is free from wind effects at 20\,\%\,\zsun{} \citep{heap06, bouret13}, and our \hst{}/COS spectra do not show emission or blueshifted absorption in this feature.
Thus, we can safely assume that this is a photospheric line suitable for constraining \vsini{} of our metal-poor target stars.

The unbroadened theoretical spectra we use in this analysis are \tlusty{} \citep{hubeny95} models drawn from the \textsc{ostar2002} grid\footnote{\url{http://tlusty.oca.eu/Tlusty2002/tlusty-frames-OS02.html}} \citep{lanz03}.
These model atmospheres span a wide range of $Z$ ($0 - 2$\,\zsun{}, with $Z_\odot=0.017$; \citealt{grevesse98}) and include NLTE and line-blanketing effects, but not stellar winds. 
At each metallicity, theoretical spectra are provided for $27.5 \leq T_\mathrm{eff} \leq 55\,\mathrm{kK}$ in steps of 2.5\,kK and for $3.0 \leq \log(g) \leq 4.75$ in steps of 0.25. 

For each star, we adopt the \textsc{ostar2002} model with \trans{c}{iii}{1176} EW matched to that measured in its COS spectrum (Table~\ref{tab:ews}).
The EW is sensitive to many stellar parameters, including \teff{}, \logg{}, and $Z$, but the intrinsic profile shape is not sensitive to these properties.
We will constrain the stellar parameters from SED fitting in Section~\ref{sec:sed_fitting} below, but for now use a simplified selection of appropriate theoretical spectra that reproduce each star's \spec{c}{iii} line strength.
These \vsini{} measurements will be used in the SED fitting process to smooth the model spectra; this is necessary to obtain a good match to the FUV continuum level.
We have checked that using the \textsc{ostar2002} models corresponding to the best-fit stellar parameters from SED fitting as the base for the \vsini{} measurements instead of those selected here does not affect our results.

We fix \logg{} to 4.0 for all three stars, as expected for luminosity class V.
To approximate the $Z$ of each star based on its host galaxy nebular $Z$ (Table~\ref{tab:galaxies}), we adopt the S, V, and W (1/5, 1/30, and 1/50\,\zsun{}) grids for A15, S3, and LP26, respectively.
These are also the $Z$ grids that we find best reproduce the stellar SEDs in Section~\ref{sec:sed_fitting} below.
Finally, we search over all available \teff{} and select the model with the \trans{c}{iii}{1176} EW that most closely matches the measured EW.
This selection process resulted in using theoretical spectra with \teff{} of 40, 35, and 37.5\,kK as the base models for A15, S3, and LP26, respectively. 
These are within the uncertainties of the best-fit \teff{} from SED modeling, discussed below.

We use the IDL routine \texttt{lsf\_rotate.pro}\footnote{\url{https://idlastro.gsfc.nasa.gov/ftp/pro/astro/lsf\_rotate.pro}} (reimplemented in Python) to generate rotational broadening profiles.
A linear limb-darkening law is assumed (equation 17.11 in \citealt{gray92}) with limb-darkening coefficient $\epsilon=0.6$ constant across the disk.
We convolve the rotationally broadened model spectra with a Gaussian of 50\,km\,s$^{-1}$ FWHM to account for the COS instrumental resolution.
We use a least-squares routine to fit for the \vsini{} and stellar radial velocity (\vrad{}) that produce the best match between the broadened \tlusty{} model spectra and the observed \trans{c}{iii}{1176} profiles for the three target stars.
The best-fit \vsini{} and \vrad{} are reported in Table~\ref{tab:velocities}.
We also repeat the host galaxy systemic velocities (\vsys{}; initially reported in Table~\ref{tab:obs} above) determined from fitting the ISM lines (Section~\ref{sec:cos_spectra}) for comparison to the measured stellar \vrad{} and find only small velocity offsets between our target stars and their host galaxies, $\leq 35\,\mathrm{km\,s}^{-1}$.

Figure~\ref{fig:vsini} shows the best-fit broadened \tlusty{} \trans{c}{iii}{1176} profiles for the three stars. 
The observed spectra are shown in black and the best-fit models in blue, with the inferred \vsini{} reported in the legends in each panel, rounded to the nearest 10\,km\,s$^{-1}$.
The light grey shows the \tlusty{} models with \vsini\,$=$\,0, broadened to COS resolution. 
The highly structured \trans{c}{iii}{1176} profile in A15 indicates a lower \vsini{} than in S3 and LP26, both of which have smoother profiles with broad wings.
The high \vsini{}\,$\geq$\,290\,km\,s$^{-1}$ for the both of the lower-$Z$ targets is somewhat surprising relative to the observed typical \vsini{}\,$\lesssim 120\,\mathrm{km\,s}^{-1}$ in the Milky Way and Magellanic Clouds, though comparable and even higher \vsini{} have been measured for OB stars in all three galaxies \citep[e.g.,][]{penny09, ramirez-agudelo13, ramachandran19}.
We discuss implications of this finding in Section~\ref{sec:implications_rotation} below.

The uncertainty in \vsini{} is dominated by modeling choices rather than measurement uncertainties.
We estimate uncertainties by repeating the \vsini{} and \vrad{} measurements using different \tlusty{} base models, changing \teff{} by one step in the grid ($\pm 2500\,\mathrm{K}$), which changes the \trans{c}{iii}{1176} by about 0.2\,\AA{} (larger than the uncertainties on our EW measurements; see Table~\ref{tab:ews}).
This exercise results in changes of $\pm 90\,\mathrm{km\,s}^{-1}$ in \vsini{} for both of the fast-rotating stars LP26 and S3, so we adopt this value as a conservative estimate of the uncertainty in those measurements.
For the lower-\vsini{} A15, however, the change due to model choice is only $\pm 10\,\mathrm{km\,s}^{-1}$.
The uncertainty in \vsini{} for A15 is likely dominated by the possible contribution of other broadening mechanisms, which we discuss in the next paragraph.
We also check the impact of several modifications to the fitting method: first, including the \trans{he}{ii}{1640} line in the modeling (the only other photospheric feature whose profile is discernible by eye in all three spectra); second, allowing the continuum level to be a free parameter in the modeling; and third, using the \tlusty{} models with \teff{} and \logg{} determined from our SED fitting in Section~\ref{sec:sed_fitting} below.
All of these result in differences no larger than $\pm 20\,\mathrm{km\,s}^{-1}$ in \vsini{}.
The \vrad{} is consistent within a few km\,s$^{-1}$ across all of these tests and is thus insensitive to model choice.

A caveat is that rotation is not the only mechanism that broadens spectral lines: macroturbulence and microturbulence also contribute.
The \textsc{ostar2002} model grid was computed with a fixed microturbulent velocity of 10\,km\,s$^{-1}$, a fairly typical value for mid-late O dwarfs; this process is not expected to dominate the observed line profiles.
The relative contributions of rotation and macroturbulence can be separated by measuring \vsini{} using the Fourier transform method \citep{gray76}.
Unfortunately, the SNR and spectral resolution of our COS data are not sufficient to enable this more precise measurement of \vsini{}. 
\citet{simon-diaz14a} suggest that derived \vsini{} for O stars that do not account for macroturbulent broadening should be revised downward by 5$-$45\,km\,s$^{-1}$ for \vsini{} below 120\,km\,s$^{-1}$.
Though the mechanism behind macroturbulence is not settled, there are observational hints that macroturbulence contributes less to broadening with decreasing metallicity and in less evolved stars \citep[e.g.,][]{penny09}.
Furthermore, \citet{bouret13} found agreement within the error bars between \vsini{} measured for SMC O dwarfs using the Fourier transform technique and by fitting rotationally broadened models, similar to our analysis here.
Our measured \vsini{} should therefore be taken as upper limits, and the true \vsini{} of A15 may be significantly lower than inferred from our modeling.
We therefore adopt $\pm 45\,\mathrm{km\,s}^{-1}$ as a conservative estimate of the uncertainty on our \vsini{} measurement for A15.
But importantly, the contribution of macroturbulence would not strongly affect the higher \vsini{} measurements, so our conclusion that S3 and LP26 are fast rotators remains secure.


\begin{table*}
\caption{Best-Fit Stellar Parameters from SED Modeling}
\label{tab:sedfit}
\begin{center}
\tabcolsep=0.2cm
\begin{tabular}{lcc|ccccc|ccc}
Star & \textsc{ostar2002} & Extinction & $T_\mathrm{eff}$ & $\log(g)$ & $A_V^\mathrm{host}$ & $f_\mathrm{FUV}^\mathrm{nebular}$ & $R_\star$ & $M_\star^\mathrm{init}$ & Age & $\log(L_\star)$ \\
 & Grid, $Z/Z_\odot$ & Law & (kK) & (cm\,s$^{-2}$) & (mag) & & ($R_\odot$) & ($M_\odot$) & (Myr) & ($L_\odot$)
\\ \hline

LP26 & W, 2\% & F99 & $37.5^{+5.9}_{-5.5}$ & $4.00^{+0.17}_{-0.24}$ & $0.01^{+0.06}_{-0.01}$ & $0.19^{+0.06}_{-0.01}$ & $8.2^{+1.3}_{-0.4}$ & $22^{+7}_{-5}$ & $7^{+4}_{-3}$ & $5.1^{+0.2}_{-0.2}$ \\
S3 & V, 3\% & F99 & $32.5^{+2.9}_{-3.0}$ & $3.50^{+0.08}_{-0.15}$ & $0.15^{+0.08}_{-0.04}$ & $0.03^{+0.05}_{-0.02}$ & $12.1^{+1.5}_{-0.3}$ & $21^{+4}_{-2}$ & $9^{+1}_{-2}$ & $5.1^{+0.2}_{-0.1}$ \\
A15 & S, 20\% & G03 & $37.5^{+7.2}_{-2.7}$ & $4.00^{+0.22}_{-0.10}$ & $0.05^{+0.06}_{-0.01}$ & $0.00^{+0.16}_{-0.00}$ & $9.5^{+0.5}_{-0.9}$ & $29^{+13}_{-3}$ & $4^{+1}_{-3}$ & $5.3^{+0.2}_{-0.1}$ \\
\end{tabular}
\end{center}
\tablecomments{The \textsc{ostar2002} grid $Z$ (reported relative to $Z_\odot=0.0017$; \citealt{grevesse98}) and extinction law are chosen for each star to best reproduce the observed FUV spectral features. We adopt the \citet{fitzpatrick99} law with $R_V=3.1$ (F99) for LP26 and S3 and the \citet{gordon03} law with $R_V=2.74$ (G03) for A15. The center five columns give the best-fit parameters from our SED modeling, while the last three columns report parameters of the \parsec{} model with \rstar{} that gives the best SED normalization.}
\end{table*}

\subsection{Estimates of Stellar Parameters from SED Fitting\label{sec:sed_fitting}}

\begin{figure*}[!ht]
\begin{centering}
  \includegraphics[width=\linewidth]{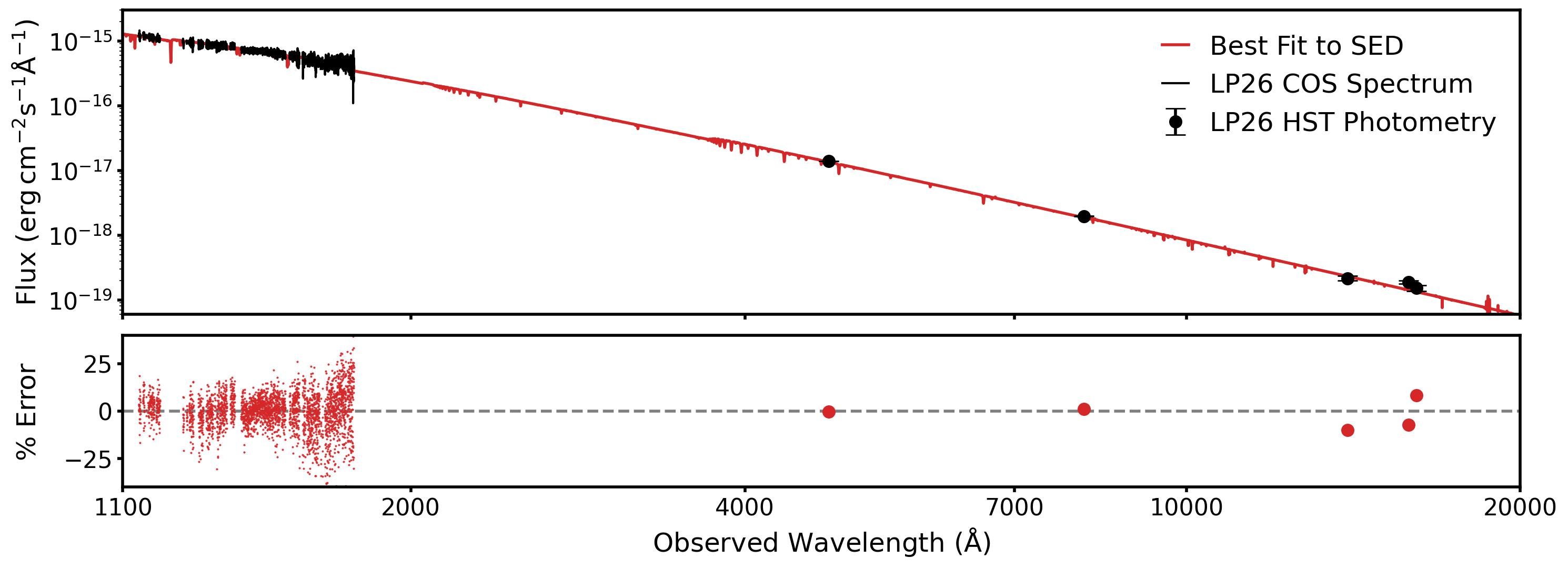}
  \includegraphics[width=\linewidth]{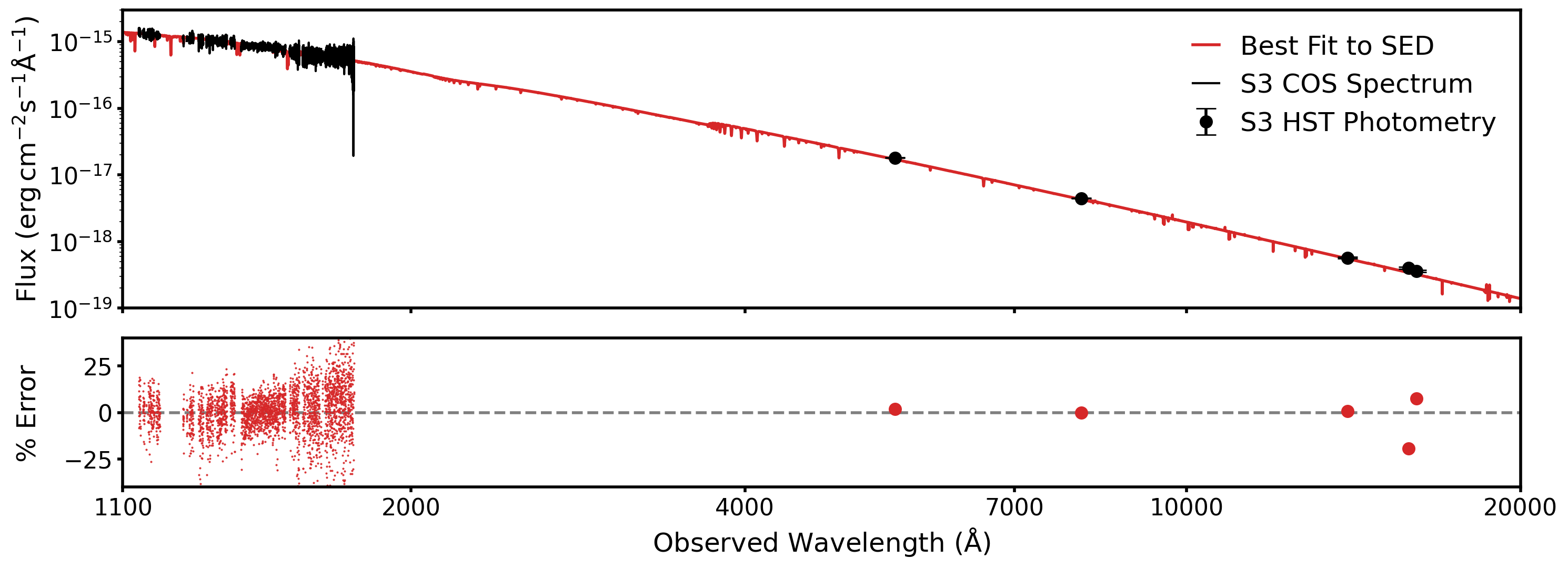}
  \includegraphics[width=\linewidth]{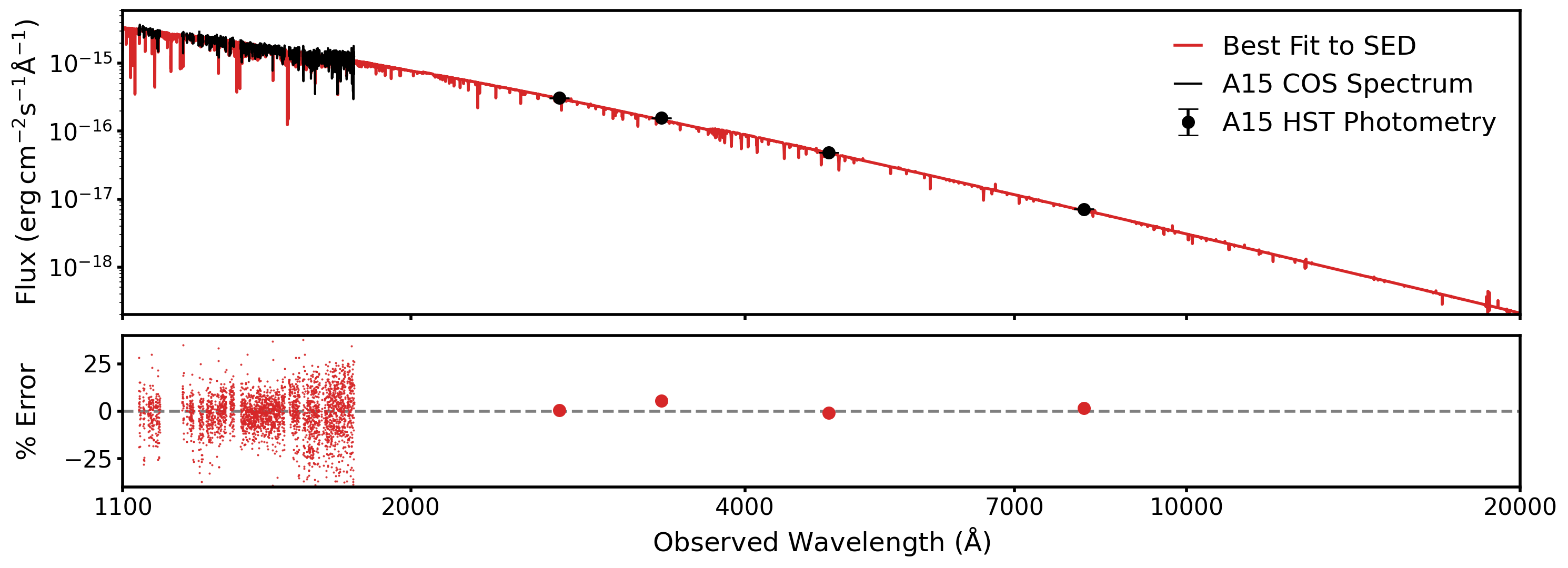}
\caption{\textbf{Models fit to the observed stellar SEDs.} From top to bottom, each pair of panels shows SED fitting results for LP26, S3, and A15. The observed FUV spectrum and \hst{} photometry (black line and points with error bars, respectively) and best-fit SED model (red line) are shown in the top panel of each pair, and the percent errors are plotted as a function of wavelength in the bottom panel, with a dashed grey line at zero error for reference. The fit quality is quite good with errors well-distributed about zero over the full wavelength range. Parameters of the best-fit models are reported in Table~\ref{tab:sedfit}. 
\label{fig:sedfit_full}}
\end{centering}
\end{figure*}

\begin{figure*}[!ht]
\begin{centering}
    \includegraphics[width=\linewidth]{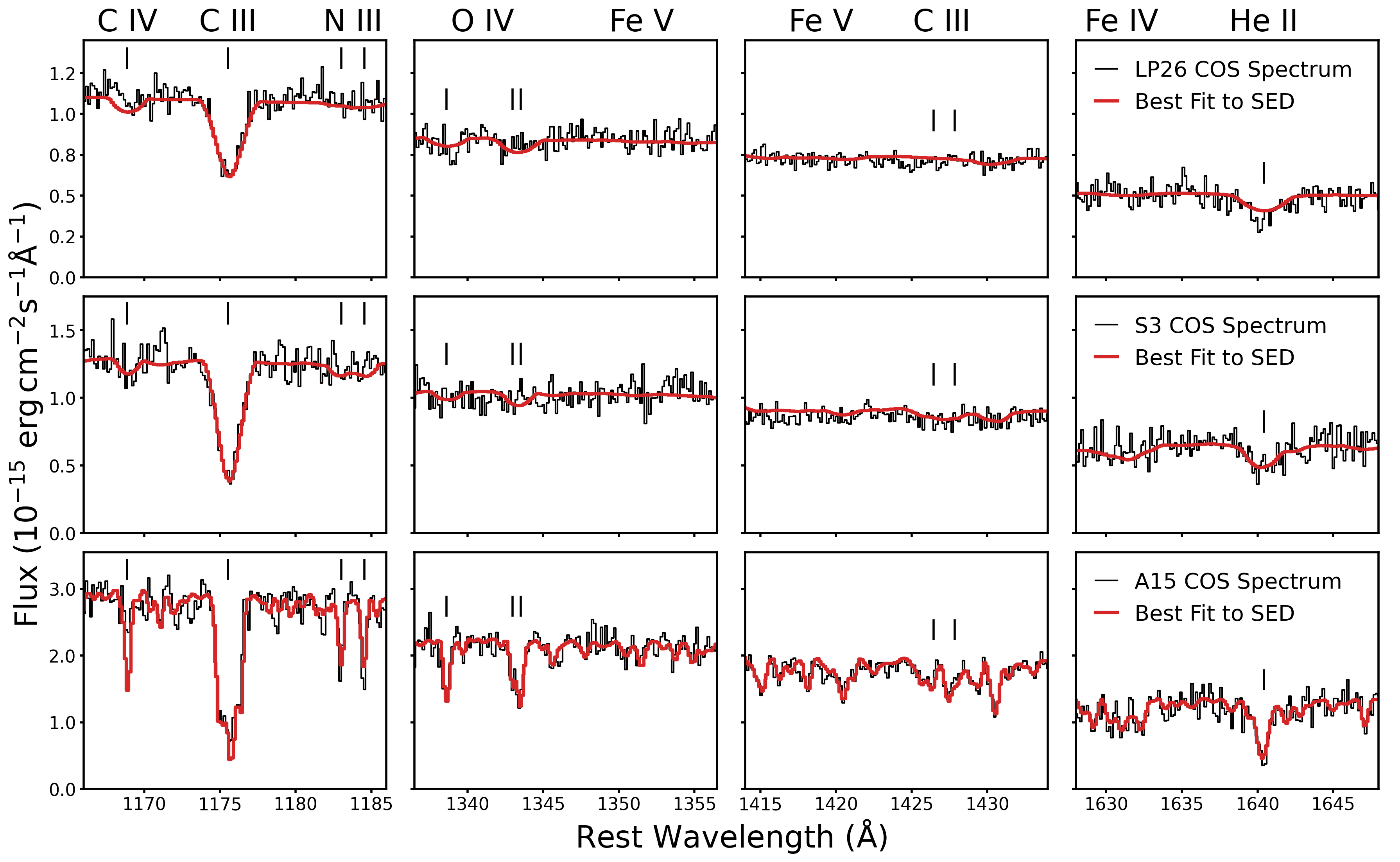}
\caption{\textbf{Comparing the best-fit SED models to observed FUV photospheric features.} From top to bottom, the observed COS spectra of LP26, S3, and A15 are shown as black lines and the best-fit models from SED fitting, convolved with the \vsini{} measured for each star (Table~\ref{tab:velocities}) and instrumental resolution, are overlaid in red. From left to right, each panel shows a small region of the spectrum focused on photospheric lines that can be used to assess consistency between the SED-based stellar parameters and the FUV properties of the stars. The wavelengths of strong photospheric lines are indicated with black vertical lines in each panel, while Fe forest regions which are simply labeled at the top of the plot. Overall, the agreement is excellent given that the model spectra are drawn from the \textsc{ostar2002} grid and the atmosphere model parameters were not tuned to match the spectral lines.
\label{fig:sedfit_fuv}}
\end{centering}
\end{figure*}

To connect the observed FUV spectral features to the physical properties of the stars, we next model their SEDs. This technique is widely used to constrain the luminosity (\lstar{}) of a star in particular, and can also estimate its \teff{}, \logg{}, initial mass (\mstar{}), and age. We simultaneously fit the continuum of the FUV spectra and the NUV, optical, and NIR photometry measured from \hst{} imaging of the three stars. All ISM, nebular emission, and wind-sensitive features are masked in the COS spectra, and the observations are corrected for Milky Way foreground extinction (Section~\ref{sec:data}). In this section, we describe our SED modeling procedure, then present the stellar properties inferred from the best-fit models and check the fit quality by comparing model-predicted and observed FUV photospheric line profiles.

\subsubsection{Description of the Modeling\label{sec:sed_models}}

We construct model SEDs using the \tlusty{} \textsc{ostar2002} atmosphere model grid (described in Section~\ref{sec:velocities} above), which provides theoretical stellar SEDs and detailed spectra as a function of \teff{}, \logg{}, and $Z$. 
The stellar $Z$ dictates the strength of photospheric metal lines as well as the \teff{} and structure of a star, such that a lower-$Z$ star at fixed \mstar{} and age will be hotter and more compact. 
Assuming a solar oxygen-to-metals ratio, the nebular abundances of the host galaxies (Table~\ref{tab:galaxies}) each fall between two of the available $Z$ values in the \textsc{ostar2002} grid, which adopts $Z_\odot=0.017$ \citep{grevesse98}.
We therefore test models at the two $Z$ values bracketing the nebular abundances for each star: the S and T grids (1/5 and 1/10\,\zsun{}) for A15; the T and V (1/30\,\zsun{}) grids for S3, and the V and W (1/50\,\zsun{}) grids for LP26.

In addition to the parameters of the star itself, dust extinction internal to the host galaxy also affects the overall shape of the stellar SED and must be accounted for in the modeling.
Dust extinction laws are known to vary widely, even among various lines of sight within the Milky Way \citep[e.g.,][]{cardelli89}.
It is therefore common to simply adopt a standard Milky Way extinction law, even when modeling stars in external galaxies, given the large uncertainty in the true shape of the extinction law.
However, the lower-$Z$ environments of the Magellanic Clouds are known to have different extinction laws, and these may be more appropriate to use when modeling stars in external, metal-poor galaxies. 
We test both a \citet{fitzpatrick99} Milky Way extinction law (F99; $R_V=3.1$) and a steeper \citet{gordon03} SMC bar extinction law (G03; $R_V=2.74$) for each star.

We repeat the SED fitting procedure described below for the four combinations of $Z$ and extinction law, and visually inspect each fit to select that which results in the best match to the continuum shape and photospheric line profiles in the FUV spectra.
The final adopted $Z$ and extinction law for each star are reported in Table~\ref{tab:sedfit}.

The \tlusty{} SED corresponding to each \teff{} and \logg{} pair is broadened to the measured \vsini{} and 50\,km\,s$^{-1}$ COS resolution and shifted to the \vrad{} of each star (Table~\ref{tab:velocities}).
The models provide the Eddington flux $H_\nu$ at the stellar surface (erg\,s$^{-1}$\,cm$^{-2}$\,Hz$^{-1}$), which must be scaled by $\left(R_\star/D\right)^2$, where \rstar{} is the stellar radius and $D$ is the distance to the star, to obtain the flux that would be measured by an observer. 
We fix $D$ to the host galaxy distance obtained from the measured magnitude of the tip of the red-giant branch (TRGB; Table~\ref{tab:galaxies}). 
The expected range in \rstar{} near each \teff{}/\logg{} grid point is obtained from the \parsec{} stellar evolution models\footnote{\url{http://stev.oapd.inaf.it/cgi-bin/cmd\_3.1}} \citep{bressan12}. 
We use isochrones spaced 0.05\,dex in $\log(\mathrm{age})$ and densely sampled in \mstar{} interpolated to the metallicities of the \textsc{ostar2002} models. 
Each sample from these evolution models has an associated \rstar{}, \mstar{}, \lstar{}, and age. 
The \rstar{} of all \parsec{} models falling within half a grid step of each \teff{} and \logg{} pair ($\pm1.25\,\mathrm{kK}$ and $\pm 0.125$\,dex, respectively) are used to scale the \tlusty{} SED at that grid point to generate a set of model SEDs with a realistic range of possible flux normalizations. 

The final ingredient in the SED models is the contribution of nebular continuum emission in the FUV.
Any nebular emission within the 2\farcs{}5 COS aperture will contribute to the observed FUV spectrum, but the nebular contribution is removed from the NUV-NIR photometry by \dolphot{}'s local background subtraction. 
This is important for LP26 in particular, as this star powers a bright \hii{} region; its NUV acquisition image inset in Figure~\ref{fig:imaging} shows some extended emission within the COS aperture. 
We use FSPS, an SPS code that self-consistently models stellar populations and the nebular emission they power \citep{conroy09, byler17}, to calculate the fractional contribution of the nebular continuum to the total (stellar + nebular) flux in the FUV. 
Over a range of stellar population $Z$, maximum \mstar{}, and age, we find that a simple polynomial model describes the shape of the fractional contribution of the nebular continuum as a function of wavelength, with normalization (i.e., the maximum fractional contribution in the FUV) \fracneb{}.
The nebular contribution is larger at redder FUV wavelengths and is negligible blueward of 1216\,\AA{}.  
To account for the nebular continuum in the SED fitting, between $1216\leq \lambda \leq 1800$\,\AA{} only, the model spectrum is multiplied by

\begin{equation}
1 + f^\mathrm{nebular}_\mathrm{FUV} \left(-6.012 + 0.008\lambda - 2.231\times10^{-6} \lambda^2 \right).
\end{equation}

For every \rstar{}, \teff{}, and \logg{} combination, we use a least-squares algorithm to optimize two free parameters: the amount of dust extinction in the host galaxy, parameterized by $A_V$, and \fracneb{}. 
We use photometry tools implemented in the \beast{} \citep{gordon16}, a Bayesian stellar SED fitting code optimized for \hst{} photometry (but not spectroscopy), to measure the flux of the scaled \tlusty{} SEDs integrated over the \hst{} filters in which we have observed photometry. 
Finally, we compute the \chisq{} goodness-of-fit metric,

\begin{equation}
\chi^2 = \sum_i^N \left( \frac{\mathrm{obs}_i - \mathrm{model}_i}{\sigma_{\mathrm{obs}, i}}\right)^2,
\end{equation}
where $N$ is the number of data points, for each \teff{}/\logg{}/\rstar{} combination with optimized $A_V$ and \fracneb{}. 
The uncertainties on the photometry are much smaller than on the spectroscopy, so using the \chisq{} metric weights the photometry more to balance the smaller number of photometric samples (4 or 5, vs. $\sim 3000$ in the COS spectra).
The model with the lowest \chisq{} is adopted as the best fit to the data.

\subsubsection{SED Fitting Results\label{sec:sed_results}}

The SED fitting procedure is performed four times for each star, with two different extinction laws and stellar $Z$ (as described in Section~\ref{sec:sed_models} above).
We visually inspect the fits to select the extinction law and $Z$ that provide the best match to the normalization and photospheric line profiles in the observed FUV spectra.
The parameters of the model that minimized \chisq{} for each star for the adopted extinction law and $Z$ are reported in Table~\ref{tab:sedfit}. 
Figure~\ref{fig:sedfit_full} presents the SED fitting results for LP26 (top), S3 (middle), and A15 (bottom). 
The top panel in each pair compares the best-fit SED model (red line) to the data (black line and points) as a function of observed wavelength. Error bars on the photometry are smaller than the point size, though noticeably larger in the NIR than in the optical and NUV. The bottom panel shows the percent error (red points), where a positive value indicates that the model under-predicts the observation. 
A horizontal line at zero error, or a perfect match to the data, is shown in the bottom panels for reference.
Figure~\ref{fig:sedfit_full} demonstrates that we were able to achieve good fits to the observations for all three targets. 
The FUV and longer-wavelength fluxes are simultaneously matched and the residuals are centered about zero and not obviously structured. 
The larger spread in percent error in the FUV is indicative of the lower SNR of the COS data relative to that of the photometry.

Figure~\ref{fig:sedfit_fuv} presents a detailed comparison between the best-fit models and photospheric features in the FUV spectra. 
The observed spectra (black) and the best-fit models (red) are plotted as a function of rest wavelength for LP26, S3, and A15, increasing in $Z$ from top to bottom. 
Model spectra have been broadened to the measured \vsini{} and COS resolution.
Each panel from left to right shows a small portion of the FUV spectrum centered on photospheric transitions sensitive to stellar properties like \teff{}, \logg{}, and metal abundances. 

Figure~\ref{fig:sedfit_fuv} demonstrates remarkably good agreement with the photospheric line profiles in the observed COS spectra, particularly given that the model spectra were constructed from a pre-computed model grid and that no stellar atmosphere parameters (e.g., individual metal abundances, microturbulence) were tuned to match the observations. 
The continuum levels are matched well, which suggests that our choice of distances, extinction laws, and parameterization of the FUV nebular continuum are sound. 
The most notable discrepancy is in the \trans{c}{iv}{1169} line in A15 (bottom left panel), which is predicted to be much stronger than observed; the cause of this mismatch is not obvious, but future detailed atmosphere modeling of this star should provide insight.
Overall, this comparison demonstrates that the \textsc{ostar2002} grid can successfully reproduce the observed photospheric features of O-dwarf stars at very low $Z$ and supports the validity of our stellar parameter inference. 
The ability of the \textsc{ostar2002} model spectra to match the FUV line profiles at such low $Z$ is an important victory, given their widespread use in SPS codes that are used to interpret metal-poor stellar populations \citep[e.g.,][]{gutkin16} and in galaxy formation models \citep[e.g.,][]{emerick19}. 

The stellar parameters inferred from the best-fit models are fairly similar across the sample: all three stars have relatively low \mstar{} ($\lesssim 30\,M_\odot$) and \teff{} ($\lesssim 40\,\mathrm{kK}$), consistent with their mid-late O spectral types. 
LP26 and A15 both appear to be midway through their main-sequence lifetimes, while the modeling suggests that S3 is more evolved and may be leaving the main sequence.
None of the stars appears to be particularly young, with best-fit ages ranging from 4 to 9\,Myr, which is not surprising, as these stars do not inhabit vigorously star-forming regions.
The models prefer very low internal dust extinction in the host galaxies, consistent with expectations for low-$Z$ dwarf irregulars. 
Only the fit to LP26 requires a large contribution of the nebular continuum in the FUV; the other two stars are fit well with \fracneb{} fixed to zero. 
This is sensible, since LP26 is powering a bright \hii{} region while the other two stars appear far from bright emission in H$\alpha$ imaging.

To estimate the uncertainty on our inferred parameters, we use the range in parameters of all models that agreed with the best-fit model within $\sim 10\%$ for \textit{both} the photometry and spectroscopy. Specifically, we required that neither \chisq{}$_\mathrm{phot}$ nor  \chisq{}$_\mathrm{spec}$ increased by more than $1.2 N$ of the minimum \chisq{}, where the number of samples $N$ is four or five for the photometry and typically $\sim 3000$ for the COS spectra.
We adopt the difference between 16th and 84th percentile values of the distribution for each parameter and the best-fit value as the uncertainties on the measurement, reported in Table~\ref{tab:sedfit}.
While the uncertainties on the SED-based \teff{} are rather large ($\pm 3-7\,\mathrm{kK}$), the inferred \teff{} for the three stars agree within the scatter with scalings between \teff{} and spectral type at low $Z$ in the literature (e.g., \citealt{doran13, ramachandran19}).

Finally, we acknowledge that neither rotation nor binary effects were accounted for in the SED analysis.
The \parsec{} stellar evolution models do not include rotation, but their flexibility in $Z$, \mstar{}, and age sampling relative to other available evolution models was essential for the modeling technique used in this work.
Rotation should only mildly affect the stellar properties associated with a star's position on the Hertzprung-Russell diagram during its main sequence lifetime \citep[e.g.,][]{brott11, groh19}, potentially causing an underestimate of stellar age by up to 1\,Myr and an overestimate of \mstar{} by a few \msun{} \citep{camacho16}.
This level of systematic uncertainty is within our conservatively estimated error budget.
It is also possible that any (or all) of our sample stars could have undetected binary companions, but the available data do not suggest that we need to account for a companion in the SED fitting.


\section{Discussion\label{sec:discussion}}


\subsection{Comparison to Analog Stars in the SMC\label{sec:smc_comparison}}

\begin{figure*}[!ht]
\begin{centering}
\includegraphics[width=\linewidth]{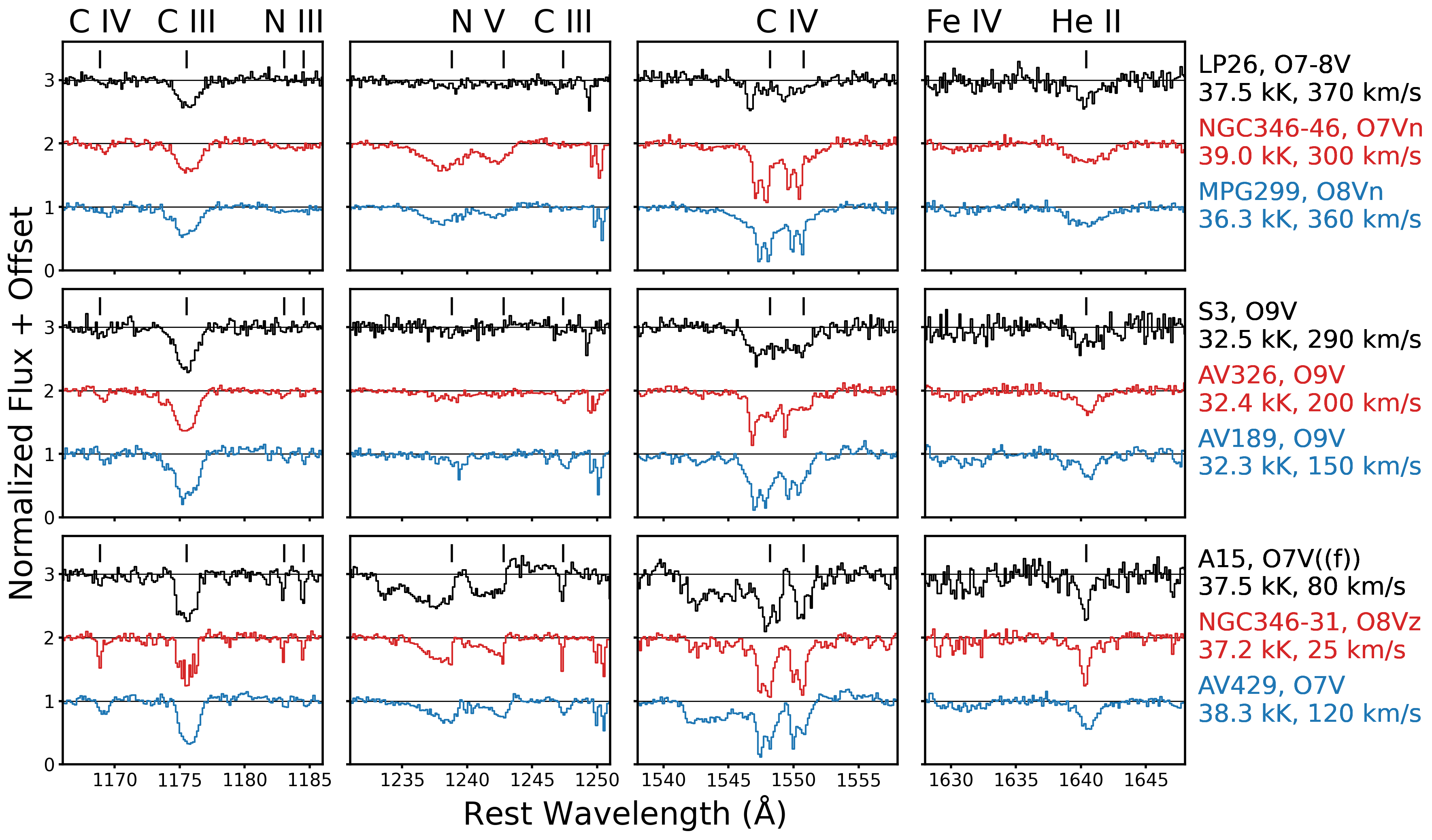}
\caption{\textbf{Comparing the FUV spectra of low-$\bm{Z}$ O dwarfs to SMC analogs.} Each row shows the continuum-normalized COS spectrum of one of our target stars in black (from top to bottom: LP26, S3, and A15) alongside the normalized spectra of two O dwarfs in the 20\%\,\zsun{} SMC in red and blue, offset for clarity. The comparison stars were matched as closely as possible in spectral type, \teff{}, and \vsini{} to the inferred parameters for our target stars (values are reported in the labels to the right of each row). Each panel shows a small region of the FUV spectrum containing photospheric and/or wind-sensitive lines, which are indicated with black vertical lines in each panel. The spectra of S3 and A15 appear surprisingly similar to their SMC analogs, but the wind lines in LP26 are far weaker than SMC stars with similar photospheric line strengths. 
\label{fig:smc_comparison}}
\end{centering}
\end{figure*}

Comparing our new FUV spectra of O~V stars in $3-14\%\,Z_\odot$ galaxies to FUV observations of similar stars in the 20\%\,\zsun{} SMC provides an empirical demonstration of the impact of $Z$ on photospheric and wind features.
The known spectral types of the target stars and measurements of their stellar properties presented in Section~\ref{sec:sed_results} enable us to identify similar stars in the SMC. 
For each low-$Z$ target, we selected two analog stars from the sample of O dwarfs in the SMC analyzed by \citet{bouret13}, for which \hst{}/COS observations are available (GO-11625; PI: I. Hubeny).
The SMC stars' fundamental properties, abundances, winds, and rotation were characterized with \cmfgen{} \citep{hillier98} atmosphere models fit to their FUV (and optical, where available) spectra. 
All stars in the \citet{bouret13} sample have \logg{} close to 4.0, but span a wide range in other properties. 
We matched the SMC stars to our targets as closely as possible in spectral type, \teff{}, and \vsini{} to control for properties other than $Z$ that affect the morphology of spectral lines.

Figure~\ref{fig:smc_comparison} compares the FUV spectra of each of our target stars to its two analogs in the SMC. 
In each row, the black line shows the spectrum of one of our targets (LP26, S3, and A15, increasing in $Z$ from top to bottom) and the \hst{}/COS spectra of two comparison stars drawn from the \citet{bouret13} sample are shown as red and blue lines. 
Each column shows a small wavelength range containing transitions sensitive to the star's photospheric conditions and/or wind strength. 
The strong wind and photospheric lines are indicated at the top of each panel with vertical black lines (except for the many \spec{fe}{iv} transitions in the right column) and labeled at the top of the figure. 
All spectra have been shifted to the star's rest wavelength and have been normalized to a continuum level of 1. 
The spectra have been offset by integer values for clarity and the horizontal lines show the continuum level of each spectrum for reference. 
Labels at the right side of the figure give the name of each star and its properties that we used to construct this comparison. 

We can now use these matched samples to assess the impact of metallicity on the FUV spectral morphology for otherwise similar O~V stars. 
Beginning with LP26 (3\%\,\zsun{}) in the top row, the photospheric features in the left and right panels appear similar across the three stars, including the smooth and broad profiles due to high \vsini{} (Section~\ref{sec:velocities}). 
This comparison to fast rotators in the SMC emphasizes the fact that high \vsini{} can broaden weak photospheric lines to the extent that they become undetectable above the noise level in the continuum, at least partly accounting for the observed featureless continuum in LP26. 
The wind-sensitive features in the center two panels, however, reveal a strong difference in wind strength between LP26 and the two higher-metallicity SMC stars. This is qualitatively consistent with expectations from line-driven wind theory.

Turning to S3 (6\%\,\zsun{}) in the middle row, both the wind and photospheric features appear similar across the three stars. 
The measured \vsini{} for S3 is higher than that of any O9~V stars in the \citet{bouret13} sample, accounting for the somewhat broader \trans{c}{iii}{1176} and \trans{he}{ii}{1640} lines. 
The wind-sensitive features also show little difference across the three stars, in the sense that they are all very weak. 
None of the spectra shows broad wind absorption in \trans{n}{v}{1240}, and all three  \trans{c}{iv}{1550} wind profiles have a small velocity extent, suggesting that the winds become optically thin in the lower-density outer regions. 
The narrow \trans{c}{iv}{1550} absorption components due to the ISM of the host galaxies and the Milky Way appear stronger in the spectra of the SMC stars, but the broad wind absorption profiles are similar for S3 and its two SMC analogs.
These are the latest spectral types in our analysis, so the finding of winds with similarly low optical depths across more than a factor of two difference in $Z$ is not surprising.

Finally, the comparison for A15 (14\%\,\zsun{}) in the bottom row tells a rather different story: both of its strong wind lines actually appear stronger than those of the SMC stars (center two panels). 
The absorption is deeper and extends blueward to higher velocities, suggesting a higher \mdot{} than in its SMC counterparts. 
Some of the photospheric metal lines appear stronger as well (specifically \trans{n}{iii}{1183, 1184} in the left panel and the \spec{Fe}{iv} lines in the right panel), though notably the \trans{c}{iv}{1169} line is much weaker in A15 than in the SMC analogs. 
The best-fit \tlusty{} SED model also failed to reproduce that weak line (Section~\ref{sec:sed_results}), making A15 an intriguing object for further study with detailed atmosphere models. 
Overall, this comparison strongly suggests that A15 is at least as metal-rich as the SMC analog stars; we discuss this finding in the context of previous work on O stars in WLM in Section~\ref{sec:individual_stars} below.
The two SMC analogs bracket A15 in \vsini{}, and the fact that the structure of A15's \trans{c}{iii}{1176} feature is intermediate between that of the two comparison stars supports our measured \vsini{}.

\begin{figure}[!t]
\begin{centering}
\includegraphics[width=\linewidth]{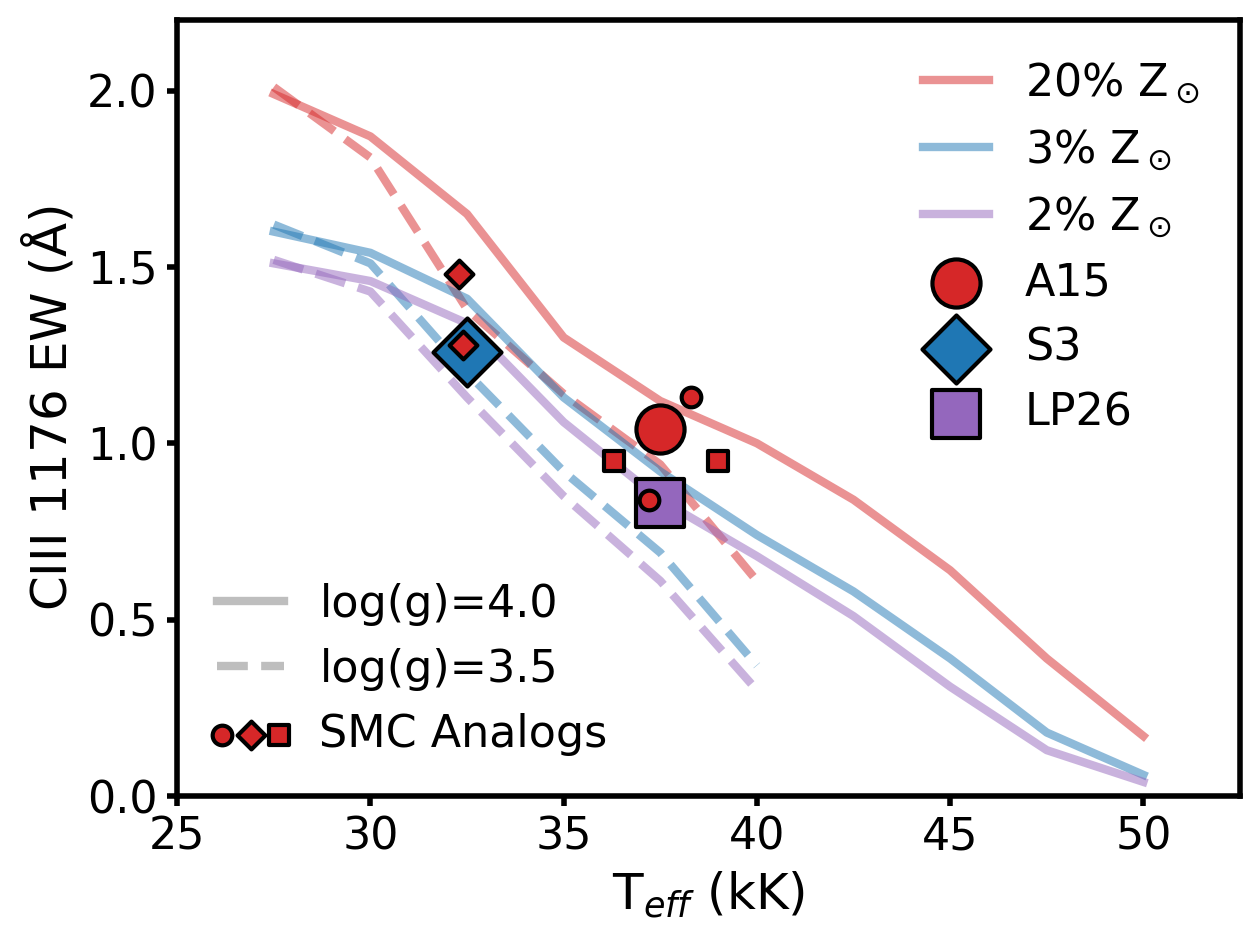}
\caption{\textbf{C\,III\,1176\,\AA{} EW in the target stars compared to their SMC analogs.} \trans{c}{iii}{1176} EW plotted as a function of \teff{} for stars in our low-$Z$ sample (large points) and their SMC analogs (small points, marker shape matched to that of their associated low-$Z$ star). Lines show \trans{c}{iii}{1176} EW measured from theoretical \tlusty{} spectra, where solid and dashed lines indicate \logg{} of 4.0 and 3.5, respectively. Color-coding indicates $Z$ of the \tlusty{} models, with red, blue, and purple corresponding to 20\%, 3\%, and 2\%\,\zsun{}, respectively. Markers for the observed stars are colored according to the $Z$ adopted in the SED fitting, or assuming 20\%\,\zsun{} for the SMC analogs. The \trans{c}{iii}{1176} EWs for LP26 and S3 are smaller than those measured in their SMC analogs, while A15's EW is intermediate between those of its SMC counterparts. Overall, \trans{c}{iii}{1176} EW is more sensitive to \teff{} than $Z$ or \logg{}, suggesting this can be a useful diagnostic of stellar population \teff{} and age in FUV observations of metal-poor galaxies.
\label{fig:ciii_ew_comparison}}
\end{centering}
\end{figure}

It is interesting that the strong photospheric lines in all three low-$Z$ target stars appear similar to those of their comparison stars in the SMC. 
As the \trans{c}{iii}{1176} line is the only one formally detected in all three of our \hst{}/COS spectra, we focus our quantitative comparison on this feature.
Figure~\ref{fig:ciii_ew_comparison} shows the measured \trans{c}{iii}{1176} EWs for our metal-poor O dwarfs (large markers, where the red circle, blue diamond, and purple square indicate A15, S3, and LP26, respectively) and for their SMC analog stars (small red markers, with shapes matched to those of the target stars against which they are compared) as a function of \teff{}.
For comparison, the lines show \trans{c}{iii}{1176} EWs measured from \tlusty{} spectra drawn from the \textsc{ostar2002} grid as a function of the model \teff{}, with solid and dashed lines indicating \logg{} of 4.0 and 3.5, respectively.
The lines are color-coded by the metallicity of the \tlusty{} models, with red, blue, and purple corresponding to 20\%, 3\%, and 2\%\,\zsun{}, respectively.
Markers for the target stars are colored to match the $Z$ of the \tlusty{} model adopted in fitting their SEDs, and SMC stars are all colored red to indicate their 20\%\,\zsun{}. 

First, the measured \trans{c}{iii}{1176} EWs agree well with the EWs predicted by their best-fit \tlusty{} models, confirming the qualitative agreement in Figure~\ref{fig:sedfit_fuv}.
Next, we compare the \trans{c}{iii}{1176} EW of each metal-poor target star to those of its SMC analogs.
Both LP26 and S3 have EWs smaller than those of their higher-$Z$ SMC counterparts at similar \teff{}, consistent with the metal-poor stars having lower C abundances. The \trans{c}{iii}{1176} EW of A15, on the other hand, is between those of its two comparison stars in the SMC, suggesting again that this star's metallicity is similar to that of the SMC.
However, the SMC stars have measured \trans{c}{iii}{1176} EWs that scatter to lower values than \tlusty{} models would predict for their reported \teff{} and \logg{} $\geq 4.0$ \citep[][]{bouret13}. 
This is perhaps unsurprising, as their properties were determined using \cmfgen{} models that fit for many parameters that are fixed in the \textsc{ostar2002} grid, like metal abundance ratios and microturbulence. 

Figure~\ref{fig:ciii_ew_comparison} demonstrates that the dependence of \trans{c}{iii}{1176} EW on \teff{} is stronger than that on \logg{} or $Z$, at least when restricted to $Z \leq 20\%$\,\zsun{}.
The EWs of the coolest target star, S3, and its SMC analogs are all offset to higher values than those of the hotter stars, despite the range in $Z$ sampled.
This raises the possibility that \trans{c}{iii}{1176} EW can be used to probe the stellar population temperature and therefore age of metal-poor galaxies. 
We suggest that \trans{c}{iii}{1176} EWs should be explored as a tool to analyze the stellar populations of low-$Z$ galaxies in the nearby universe.


\subsection{Comparison to Previous Work on the Target Stars and Their Host Galaxies\label{sec:individual_stars}}

Here, we discuss our results in the context of previous studies of the target stars and of other massive stars in the same host galaxies.
Optical spectra exist for all three stars, though their resolution and/or SNR is generally not high enough for reliable determination of stellar properties.
We compare our estimates of stellar properties with expectations based on existing data.

\paragraph{LP26} This star was identified as the lone O star in Leo~P in optical \hst{} photometry \citep{mcquinn15b} and was later confirmed to show \spec{he}{ii} absorption indicative of an O-type star in $R \sim 2000$ optical spectroscopy \citep{evans19}. 
LP26 is powering the only \hii{} region in the galaxy, and spectroscopic observations therefore sample both the stellar and nebular flux.
Strong nebular \spec{he}{i} emission precludes a detailed analysis of the optical spectrum, so its spectral type can only be estimated.
This is made difficult by the lack of an empirical calibration between bolometric correction and spectral type at very low $Z$; the fraction of the \lstar{} of a given O type emitted in the optical should decrease with decreasing metallicity, as \teff{} shifts higher for fixed spectral type.
\citet{mcquinn15b} argued based on expectations for Galactic O stars that the absolute $V$-band magnitude of LP26 suggests an O5~V type, or possibly two blended O7-8~V stars each with $M_\star \sim 25\,M_\odot$.
Our SED analysis here prefers a $22\,M_\odot$ star with $T_\mathrm{eff} \sim 37.5\,\mathrm{kK}$, the latter being consistent with O7~V stars in the SMC \citep{ramachandran19}.
We find no indications of a binary companion and do not model the star as an unresolved binary, yet our results agree with previous estimates of LP26's properties based on the assumption that it is an unresolved binary.
This can potentially be explained by accounting for the very low $Z$ of LP26 in our SED modeling, and/or the inclusion of FUV and NIR data in this analysis providing more information than the optical photometry alone.

\paragraph{S3} \citet{garcia19} identified four O-dwarf stars in the outskirts of Sextans~A, including S3 (and two ULLYSES Targets, S2 and S4). 
They used $R \sim 1000$ spectra to assign the spectral type O9~V and compared S3's extinction-corrected position on a ground-based optical CMD with stellar evolution tracks from \citet{lejeune01}. 
That comparison suggested $M_\star \sim 25-40\,M_\odot$ and an age of $\sim 1.5-6.3\,\mathrm{Myr}$. 
The present analysis assumes different distance, foreground extinction, and adopted stellar evolution models, yet paints a remarkably consistent picture given these many uncertain factors.
Our SED modeling, including FUV and NIR constraints, prefers a slightly lower-mass and more evolved star ($M_\star \sim 21\,M_\odot$, age $\sim 9\mathrm{\,Myr}$), but roughly agrees within the uncertainties reported here and in \citet{garcia19}.
The best-fit SED model agrees with the FUV photospheric features (Figure~\ref{fig:sedfit_fuv}), but many informative diagnostic lines are undetected due to relatively low SNR combined with high \vsini{}.

\citet{garcia17} presented a low-resolution (${R \sim 2600}$) \hst{}/COS spectrum of an O7.5~III star in Sextans~A.
Compared to similar spectral type stars in IC~1613 and the SMC, the photospheric metal lines and wind features appear substantially weaker in the Sextans~A star, indicating a substantially lower metal content than 20\%\,\zsun{}.
This is consistent with our SED analysis, which prefers a low $Z \sim 0.00056$ (1/30\,\zsun{} in the \textsc{ostar2002} grid, or 4\%\,\zsun{} relative to $Z_\odot = 0.0142$; \citealt{asplund09}).
O stars in Sextans~A are thus important targets for future observational efforts to constrain the astrophysics of stars more metal-poor than the SMC.

\paragraph{A15} \citet{bresolin06} presented the $R \sim 900$ optical spectrum and determined the O7~V((f)) spectral type for this star, but deemed the SNR of the data too low for parameter inference. 
No estimates of its properties exist in the literature to the best of our knowledge.

Another O-type star (A11, O9.7~Ia) in WLM has been analyzed in detail by \citet{tramper14} and \citet{bouret15}. The latter authors obtained \hst{}/COS FUV data to complement its optical spectrum and fit \cmfgen{} models to infer wind and stellar properties, including detailed abundances. 
They found an SMC-like stellar Fe abundance of 20\% solar, though could not rule out a lower value of 14\% solar given the SNR of the data.
Based on our SED analysis of A15 and comparison to analog stars in the SMC, we argue that A15 may also have SMC-like metal abundances, consistent with the only other O star in WLM observed in the FUV.
Together, these results suggest that WLM, like IC~1613 \citep{garcia14, bouret15}, is not substantially more metal-poor (if at all) than the SMC, and may have subsolar $\alpha$/Fe.


\subsection{Implications for Low-$Z$ Stellar Populations\label{sec:implications}}

We have analyzed the first FUV spectra of mid-late O~V stars in metal-poor galaxies.
Though a sample of three is obviously small, these data represent an important step toward empirical constraints on the purely theoretical spectral and evolution models that are used to model stellar populations at low $Z$.
Here we compare our findings to the theoretical expectations that underlie the current generation of SPS models.

\subsubsection{Stellar Winds\label{sec:implications_winds}}

Radiation-driven mass loss is theoretically expected to weaken with decreasing metallicity due to the lower opacity of the photospheric metal lines that transfer momentum to stellar winds \citep[e.g.,][]{castor75, abbott82, vink01}.
While this general expectation has been validated with observations down to the 20\%\,\zsun{} population of massive stars in the SMC, the precise scaling of \mdot{} with $Z$ remains unclear \citep{mokiem07, ramachandran19, bjorklund21}.
The picture becomes even murkier at lower $Z$ due to the lack of spectroscopic observations of individual metal-poor O stars in galaxies more distant than the Magellanic Clouds. 
\citet{tramper11, tramper14} reported higher \mdot{} than expected from radiation-driven wind theory based on analysis of optical spectra of several O stars in the metal-poor galaxies WLM, IC~1613, and NGC~3109, but later modeling of the wind profiles in FUV spectra of three of those stars by \citet{bouret15} found lower \mdot{}.
Furthermore, the O stars in two of those galaxies with gas-phase oxygen abundances below 20\%\,\zsun{} appear to have SMC-like Fe abundances \citep{garcia14, bouret15}, alleviating the apparent tension between observations and theory. 
The challenge of identifying stars with Fe abundances below 20\%\,\zsun{}, even in galaxies with nebular oxygen abundances below that threshold, remains a limiting factor in our understanding of stellar winds at low $Z$. 

More distant dwarf irregulars with gas-phase oxygen abundances below 10\%\,\zsun{} are the new observational frontier. 
Ten O stars have been spectroscopically confirmed in the very metal-poor galaxies Sextans~A (6\%\,\zsun{}; \citealt{camacho16, garcia19}) and Leo~P (3\%\,\zsun{}; \citealt{evans19}), but their FUV properties have yet to be analyzed in detail.
The low-resolution COS spectrum of an O giant in Sextans~A presented by \citet{garcia17} revealed weak metal absorption and wind features relative to similar stars in IC~1613 and the SMC, suggesting that it is substantially more metal-poor than 20\%\,\zsun{}. 
Our analysis of S3 in Sextans~A and LP26 in Leo~P agrees with this picture. 
Their FUV spectra show weak photospheric metal lines, particularly in Fe forest regions, though this appears to be partly due to high \vsini{} broadening the features to the extent that they cannot be detected over the noise level of the continuum.  
The SED modeling in Section~\ref{sec:sed_fitting} prefers $Z < 5\%\,Z_\odot$ for both LP26 and S3, and the best-fit models reproduce the photospheric absorption features in their FUV spectra remarkably well.
These results, combined with the comparison to SMC stars in Figure~\ref{fig:smc_comparison}, indicate that these stars are indeed significantly more metal-poor than the SMC, opening a new window onto the astrophysics of massive stars at extremely low $Z$.

Qualitatively, the FUV spectra of S3 and LP26 are consistent with recent theoretical work at 1/30\,\zsun{}. 
\citet{martins21} predict that low mass-loss winds driven by extremely low-$Z$ stars on the main sequence result in weak wind profiles that only show absorption, using the \citet{vink01} mass-loss prescription with reduced normalization to account for the effects of clumping.
Observations agree with their predicted spectra: no wind emission or well-developed P-Cygni profiles are seen in our \hst{}/COS spectra of S3 and LP26 (Section~\ref{sec:windlines}), or in the spectrum of the O7.5~III Sextans~A star presented in \citet{garcia17}.
Detailed modeling of the photospheric and wind transitions to constrain the abundances and \mdot{} of these extremely metal-poor stars is beyond the scope of this paper, but such analysis and comparison to quantitative mass-loss prescriptions in the literature will be the subject of future work.

The behavior of \mdot{} at low $Z$ has important implications for ionizing photon production, both in nearby dwarf galaxies and in the early universe.
Not only does the wind strength affect a star's lifetime and surface properties, it also determines the opacity to ionizing photons, particularly those capable of ionizing \heii{} \citep{schaerer97, martins21}.
However, different modeling approaches find different \mdot{} normalization and scaling with metallicity \citep[e.g.,][]{vink01, krticka17, bestenlehner20, vink21}.
Since most widely-used stellar evolution and SPS models adopt the same \citet{vink01} mass-loss prescription, which seems to over-predict \mdot{} relative to empirical studies in the SMC \citep{ramachandran19, bjorklund21}, it is possible that stellar evolution is being modeled incorrectly at low $Z$.
Future work to constrain \mdot{} for metal-poor stars, including new observations to expand the current small sample, is urgently needed to anchor models of low-$Z$ stellar populations.
This could help to alleviate the discrepancy between current SPS models and the hard ionizing spectra implied by nebular emission in nearby dwarf galaxies \citep[e.g.,][]{berg19, senchyna19} and will help to determine the dominant ionizing photon producers during cosmic reionization.

It is interesting that only one of the three stars in our sample, LP26, powers a strong \hii{} region.
This star does not appear to be substantially hotter or more massive than the others, yet it is the most metal-poor and has the weakest wind features of the three.
The appearance of the \hii{} region in Leo~P could be related to its low star formation rate ($2.1\times10^{-5}\,M_\odot\,\mathrm{yr}^{-1}$; \citealt{mcquinn15b}): if few massive stars capable of clearing out neutral gas have formed close to LP26 in the recent past, we would expect a high local density of neutral gas to be available for LP26 to ionize.
Yet the fact that LP26 powers strong nebular emission despite its modest inferred \teff{} of 37.5\,kK suggests that this metal-poor star with weak winds has a relatively large ionizing flux.
In future work, we will use the \hii{} region emission powered by LP26 to constrain its ionizing spectrum and shed light on whether weak winds can lead to the production of enough energetic photons to explain the extreme nebular emission seen in very low-$Z$ galaxies.

\subsubsection{Rotation\label{sec:implications_rotation}}

\begin{figure*}[!ht]
\begin{centering}
\includegraphics[width=0.75\linewidth]{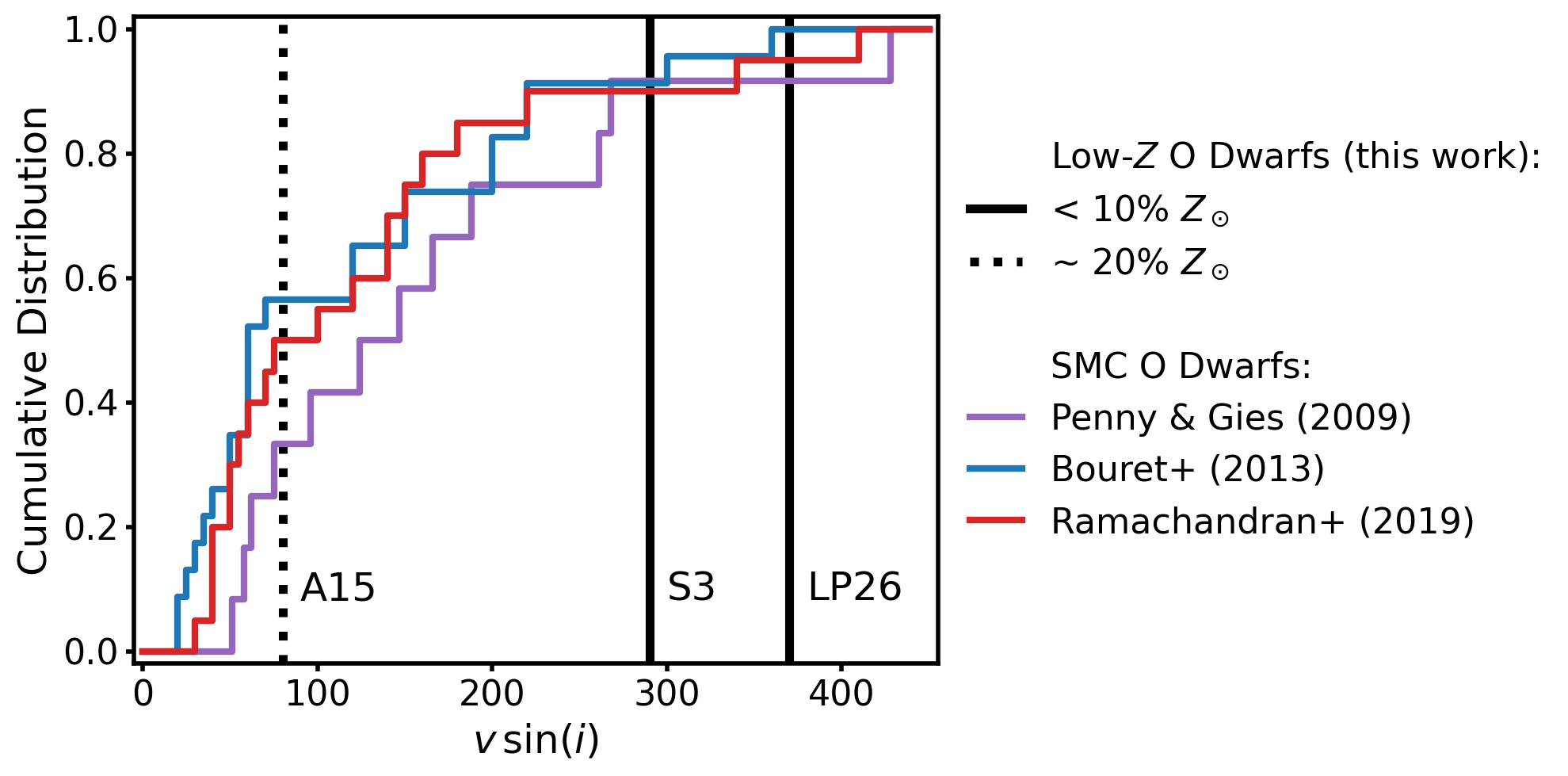}
\caption{\textbf{Both extremely metal-poor stars have $\bm{v\,\sin{(i)}}$ similar to the fastest-rotating SMC O dwarfs.}  The cumulative distribution of \vsini{} measured for three samples of O dwarfs in the SMC using different data and methods are plotted as colored lines \citep{penny09, bouret13, ramachandran19}. The \vsini{} we measure for our three metal-poor O dwarfs are indicated with black vertical lines. A15, which seems to have similar $Z$ to SMC stars, is shown as a dashed line, while the two stars that appear to be substantially metal-poorer, LP26 and S3, are shown as solid lines. While the \vsini{} of A15 falls near the median of these distributions (cumulative probability of 35-60\%), both LP26 and S3 have \vsini{} higher than 90\% of the measurements for SMC O dwarfs. Given that these were not selected to be fast rotators, this comparison suggests that the rotation speeds of very metal-poor O stars are different than the distribution observed in the SMC.
\label{fig:vsini_comparison}}
\end{centering}
\end{figure*}

Stellar rotation affects the surface properties, chemical mixing, lifetimes, and ionizing photon production of massive stars.
A wide range of \vsini{} is observed, and stars do not maintain the same rotation speed throughout their evolution.
Angular momentum loss via stellar winds is expected to slow a massive star's rotation over its lifetime, though other mechanisms like magnetic braking may also play a role.
As mass loss is expected to weaken with decreasing metallicity, winds should become less efficient at reducing surface rotation in metal-poor stars. 

Observations of large samples of OB stars in the Milky Way and Magellanic Clouds have probed the effects of  stellar age, evolutionary status, $Z$, and environment on the distribution of \vsini{}.
Across all three galaxies, OB stars appear to follow a bimodal \vsini{} distribution with most stars populating a relatively low-velocity peak and a smaller fraction forming a long tail to high \vsini{}, reaching up to $\sim 600\,\mathrm{km\,s}^{-1}$ in the LMC \citep[e.g.,][]{ramirez-agudelo13, ramachandran19}. 
The tail of fast rotators is thought to be produced by binary interactions, with secondaries spun up by mass transfer, which may be more efficient at low $Z$ \citep{de-mink13}.
Interestingly, a recent compilation by \citet{ramachandran19} showed that the peak of the low-velocity part of the \vsini{} distributions shifts higher with decreasing metallicity, as does the fraction of OB stars populating the high-velocity tail ($v\sin(i) \geq 200\,\mathrm{km\,s}^{-1}$).

We found that the two lower-$Z$ O dwarfs in our sample, both in galaxies with nebular oxygen abundances $\leq$6\%\,\zsun{}, have \vsini{}\,$\geq$\,290\,km\,s$^{-1}$ (Section~\ref{sec:velocities}). 
These are the highest reported \vsini{} of O stars in galaxies more metal-poor than the SMC, which is striking because rotation or line widths were not used in sample selection. 
Broadening measurements, giving an upper limit on \vsini{}, were reported for a handful of O stars in WLM, IC~1613, and NGC~3109, spanning a fairly broad distribution with typical values $\sim 100\,\mathrm{km\,s}^{-1}$ \citep{tramper11, tramper14, garcia14}. 
However, none of these galaxies has been demonstrated to host O stars that have lower Fe abundances than 20\%\,\zsun{}.
The only other \vsini{} measurements for stars that are securely below SMC metallicity were recently reported by \citet{ramachandran21}, who analyzed three O stars in the $\sim 10\%\,Z_\odot$ Magellanic Bridge, tidal debris from past interaction between the LMC and SMC.
Curiously, two out of the three stars in that sample were found to be extremely narrow-lined slow rotators, with \vsini{} less than 10\,km\,s$^{-1}$.
Both stars appear to have binary companions, and the authors suggest the slow rotation may be due to tidal locking.
Our \vsini{} measurements for LP26 and S3 appear to be the first constraints on how stellar rotation operates in (apparently) single extremely metal-poor O stars.
While it is possible that our target stars have binary companions, the data are entirely consistent with them being single stars and companions are not required to explain our observations.

Figure~\ref{fig:vsini_comparison} compares the \vsini{} we measure for our metal-poor sample to the distribution of \vsini{} observed for O dwarfs in the SMC.
The colored lines show three cumulative distributions of \vsini{} measured for SMC O dwarfs (defined as luminosity class IV and V here) using different samples, data, and measurement techniques \citep{penny09, bouret13, ramachandran19}.
These samples contain between 12 and 23 stars.
We indicate the \vsini{} measured for our three target stars as vertical black lines, where A15 is shown as a dotted line to clarify that it appears to be similar in $Z$ to the SMC sample, while the substantially more metal-poor LP26 and S3 are shown as solid lines.
The fraction of stars in the SMC distributions that have lower \vsini{} than a given value can be read off from this plot as the y-axis value corresponding to that \vsini{}.
A15's \vsini{} is larger than that of $\sim35-60$\% of the SMC stars, placing it close to the median of the \vsini{} distribution for SMC O dwarfs.
In contrast, both LP26 and S3 have \vsini{} that is higher than 90\% of the SMC distribution.
This is strongly suggestive that the rotation speeds of extremely metal-poor O dwarfs ($< 10\,\%$\,\zsun{}) are drawn from a different distribution than that of similar stars in the SMC, in the sense that very metal-poor stars may typically rotate faster.

How likely would it be to find that two out of our three metal-poor stars are fast rotators if their \vsini{} distribution is the same as that of SMC O dwarfs?
Taking 290\,km\,s$^{-1}$ (the \vsini{} of S3) as the lower limit, Figure~\ref{fig:vsini_comparison} shows there is a 10\% chance that SMC stars rotate faster than that.
We can use the binomial distribution to calculate the probability of successes in two out of three trials given a 10\% chance of success on independent trials:

\begin{equation}
\mathrm{Pr}\left(k \,|\, n, p\right) = \frac{n!}{k! \left(n-k\right)!} \, p^k \left(1 - p\right)^{n-k},
\end{equation}
where $k=2$, $n=3$, and $p=0.1$.
This gives just a 3\% chance of finding such high \vsini{} for two of our targets if very metal-poor stars do have a similar \vsini{} distribution to that in the SMC. 
However, we have argued that A15 has a relatively high $Z$ similar to SMC stars, so it may not be fair to count that star as being metal-poor.
We repeat the above exercise with the following modifications: (1) we adopt the lowest \vsini{} allowed within the uncertainties for the two very metal-poor stars, 200\,km\,s$^{-1}$ (Section~\ref{sec:velocities}), as a more conservative threshold, and (2) we calculate the probability of finding that two out of two very metal-poor O stars are fast rotators by this new definition.
200\,km\,s$^{-1}$ is larger than $\sim75-85$\% of the \vsini{} measured for SMC O dwarfs, so we adopt a conservatively high 25\% chance of success per trial.
With $k=2$, $n=2$, and $p=0.25$, we find a 6\% probability that metal-poor O dwarfs have the same \vsini{} distribution as their SMC counterparts.
While our sample size is obviously small, this exercise strongly suggests that extremely metal-poor O stars typically rotate faster than higher-$Z$ stars in the SMC.

It is interesting to compare these high \vsini{} to the rotation speeds at which the stars would break apart.
Using the \mstar{} and \rstar{} inferred from the best-fit SED models for each star (Section~\ref{sec:sed_fitting}), we can estimate their critical rotation speeds as:

\begin{equation}
v_\mathrm{crit} = \left( \frac{GM_\star}{R_{\star}}\right) ^{1/2}.
\end{equation}
For simplicity, we use the \rstar{} inferred from non-rotating evolution models, but in detail, fast-rotating stars are nonspherical. The equatorial radius is 1.5 times the polar radius at critical rotation, which would drive $v_\mathrm{crit}$ lower, so our simple calculation gives a conservatively high estimate of $v_\mathrm{crit}$. We find that \vsini{} measured for LP26 and S3 are 52\% and 50\% of their $v_\mathrm{crit}$, comparable to but slightly higher than the $40\% \, v_\mathrm{crit}$ often adopted in rotating evolutionary models (e.g., Geneva and MIST; \citealt{ekstrom12, choi16}). A15's measured \vsini{} is about $10\% \, v_\mathrm{crit}$.

The best-fit SED models favor ages of at least several Myr for all three stars in our sample, suggesting that all of these low-$Z$ O dwarfs have progressed through a large part of their main-sequence lifetimes.
S3 appears to be the most evolved of the three, with a best-fit age of 9\,Myr. 
The high \vsini{} of LP26 and S3 suggest that either these stars have maintained high rotation rates throughout a large fraction of their main-sequence lifetimes, or potentially could be the spun-up remnants of mass transfer from a former binary companion \citep{de-mink13}. 
Neither star shows strong wind absorption profiles (Section~\ref{sec:windlines}), so their stellar winds must be optically thin and therefore removing little mass and angular momentum from the stars.
Our findings are consistent with the expectation that metal-poor massive stars spin down less than their higher-$Z$ counterparts \citep[e.g.,][]{meynet02, brott11, groh19}, though it is impossible to differentiate between this possibility and potential binary interactions.

Trends between \vsini{} of OB stars and their environments may provide a clue.
Specifically, the rapidly rotating tail of the \vsini{} distribution is more populated in the field of the LMC compared to clusters \citep{ramirez-agudelo13}. 
The reason for this is not clear, but could be explained if spun-up secondaries are frequently ejected from their natal clusters by the supernovae of the shorter-lived primaries.
In such a scenario, we would expect to find a correlation between the \vsini{} of field stars and large radial velocity offsets from the galaxy systemic velocity.
As reported in Table~\ref{tab:velocities}, we find no large \vrad{} discrepancies between our target stars and their host galaxies that would suggest they had been kicked by a companion star's supernova. 

Our findings suggest that rotation is important to include in stellar evolution and SPS models at low $Z$.
Rotation can boost ionizing photon production by massive stars \citep[e.g.,][]{szecsi15, topping15, murphy21}, and so is important to include in modeling of the metal-poor stellar populations that power nebular emission in local dwarf galaxies and contributed to the reionization of the universe.
But the small number of \vsini{} measurements at low $Z$ limits our ability to draw broad conclusions.
Future observational efforts should aim to obtain moderate-resolution spectra of more O-type stars in nearby, metal-poor galaxies from which \vsini{} can be constrained.
Only with a statistical sample spanning a range of spectral types and luminosity classes can we determine whether the apparent trend for lower-$Z$ stars to maintain higher rotation speeds over their main-sequence lifetimes is real.


\section{Conclusions\label{sec:conclusions}}

We have presented new medium-resolution \hst{}/COS FUV spectra of the metal-poor O~V stars LP26, S3, and A15, residing in nearby dwarf galaxies spanning $3-14\%$\,\zsun{}. From these data, we measured the strengths of their photospheric and wind lines and constrained their projected rotation speeds from the structure of the \trans{c}{iii}{1176} feature. We combined the FUV spectra with NUV, optical, and NIR \hst{} photometry and modeled the SEDs to estimate key stellar parameters, including \teff{}, \logg{}, \mstar{}, \lstar{}, and age. Finally, we compared the observed FUV spectra of the metal-poor O dwarfs to those of stars with similar properties in the 20\%\,\zsun{} SMC to assess empirically the impact of $Z$ on FUV spectral properties. Our conclusions are:
\begin{enumerate}
\item Many photospheric metal lines are clearly detected in the highest-$Z$ star, A15, but the stellar continuua are strikingly featureless in the two very low-$Z$ stars, LP26 and S3. This is likely due to a combination of intrinsically weak lines and broadening due to high \vsini{}. We measure equivalent widths of the photospheric lines as a benchmark for models at low $Z$ (Section~\ref{sec:opacities}, Figure~\ref{fig:spectra}, Table~\ref{tab:ews}). 
\item The most metal-poor star, LP26, has no clear wind features, while S3 shows under-developed \trans{c}{iv}{1550} wind absorption that likely arises only in the inner, denser part of the wind. A15 has two well-developed (though unstaurated) wind profiles that suggest $v_\infty \gtrsim 1400\,\mathrm{km\,s}^{-1}$, similar to measurements for mid-late O stars in the SMC (Section~\ref{sec:windlines}, Figure~\ref{fig:windlines}, Table~\ref{tab:windlines}).
\item The two extremely low-$Z$ stars, LP26 and S3, have \vsini{}\,$\geq$\,290\,km\,s$^{-1}$, supporting the expectation that angular momentum loss via stellar winds becomes less efficient with decreasing metallicity. The high rotation speeds broaden weak photospheric lines such that they cannot be detected above the noise in the FUV spectra. A15, on the other hand, has a relatively low \vsini{} of 80\,km\,s$^{-1}$ (Section~\ref{sec:velocities}, Figure~\ref{fig:vsini}, Table~\ref{tab:velocities}). 
\item We estimate just a $3-6\%$ probability of finding that both LP26 and S3 have such high \vsini{} if they are drawn from the same \vsini{} distribution as observed in the SMC. This strongly suggests that we should expect the population of extremely metal-poor massive stars ($<10$\%\,\zsun{}) to have higher typical rotation speeds than more metal-rich populations (Section~\ref{sec:implications_rotation}, Figure~\ref{fig:vsini_comparison}).
\item The best-fit SED models suggest that all three of our targets are relatively low-mass ($M_\star \lesssim 30\, M_\odot$) O stars that have evolved through a substantial fraction of their main-sequence lifetimes (ages of $\sim 4-9\,\mathrm{Myr}$; Section~\ref{sec:sed_fitting}, Figure~\ref{fig:sedfit_full}, Table~\ref{tab:sedfit}). We compare the the best-fit model predictions of photospheric line profiles and continuum levels in the FUV to the observed COS spectra and find excellent agreement overall, supporting our inferred stellar properties and validating the use of the theoretical \textsc{ostar2002} atmosphere grid down to very low $Z$ (Figure~\ref{fig:sedfit_fuv}).
\item We compare FUV spectra of the three metal-poor O stars to analogs in the SMC, controlling for the effects of all stellar properties except $Z$. A15 shows photospheric and wind features at least as strong as those in its SMC analog stars, suggesting that its stellar $Z$ may be higher than the gas-phase oxygen abundance of WLM. Only the most metal-poor star, LP26, has obviously weaker wind lines than its SMC analogs, supporting the expectation of lower \mdot{} with decreasing metallicity (Section~\ref{sec:smc_comparison}, Figure~\ref{fig:smc_comparison}). Both LP26 and S3 have smaller \trans{c}{iii}{1176} EWs than their SMC analogs, while A15's is intermediate between its comparison stars. We suggest that \trans{c}{iii}{1176} EW may be a useful age diagnostic for unresolved massive stellar populations at low $Z$ (Figure~\ref{fig:ciii_ew_comparison}).
\end{enumerate} 

These results provide a new window into the astrophysics of extremely metal-poor O-dwarf stars.
Though the sample size is small, this work demonstrates that FUV spectra of individual massive stars contain a wealth of information and can be used to anchor theoretical stellar evolution and atmosphere models upon which our understanding of the ionizing flux and feedback of metal-poor stellar populations depend.
In the future, we will fit detailed stellar atmosphere models to the FUV and optical spectra of these metal-poor O dwarfs to quantify the abundances of individual metals and constrain the wind properties, particularly \mdot{} and \vinf{}.
The inferred stellar and wind properties will enable a unique test of mass-loss models below 20\%\,\zsun{}, and the atmosphere models will enable a data-driven estimate of the ionizing flux from metal-poor O stars.


\acknowledgements

OGT thanks Ben Williams and Karl Gordon for helpful discussions and Max Newman for contributions to the photometry pipeline used in this work.

This work was supported by the Space Telescope Science Institute through GO-15967. 
OGT was also supported by Rutgers University.
This work is based on observations made with the NASA/ESA \textit{Hubble Space Telescope}, obtained from the data archive at the Space Telescope Science Institute. 
STScI is operated by the Association of Universities for Research in Astronomy, Inc. under NASA contract NAS 5-26555. 
This work was performed in part at Aspen Center for Physics, which is supported by National Science Foundation grant PHY-1607611. 
This work was partially supported by a grant from the Simons Foundation.

This research has extensively used NASA's Astrophysics Data System, adstex\footnote{\url{https://github.com/yymao/adstex}}, and the arXiv preprint server. 
The Digitized Sky Survey was produced at the Space Telescope Science Institute under U.S. Government grant NAG W-2166. 
The images of these surveys are based on photographic data obtained using the Oschin Schmidt Telescope on Palomar Mountain and the UK Schmidt Telescope. 
The plates were processed into the present compressed digital form with the permission of these institutions.

\software{Astrodrizzle \citep{gonzaga12}, Astropy \citep{astropy, astropy2}, \beast{} \citep{gordon16}, \dolphot{} \citep{dolphin00, dolphin16}, HDF5 \citep{hdf5}, iPython \citep{ipython}, Matplotlib \citep{matplotlib}, NumPy \citep{numpy2}, SAOImageDS9 \citep{ds9}, SciPy \citep{scipy2}}


\appendix
\section{Continuum Normalization}\label{sec:appendix}
\setcounter{table}{0}
\renewcommand{\thetable}{A\arabic{table}}

Table~\ref{tab:contnorm} reports the spectral bandpasses used to fit the local continuum level for each spectral line analyzed in this paper. We selected a narrow range of wavelengths on either side of the spectral line, ensuring that the continuum regions were free of both ISM lines and strong photospheric lines. To ensure that we weren't missing photospheric absorption masked by the noise in the continuum, we used \tlusty{} spectra to check for predicted metal absorption in all of the bandpasses we used. For most lines, we modeled the local continuum with a simple linear fit to the spectrum in the blue and red bandpasses, and visually inspected all fits and normalized spectra to confirm that this procedure produced good results. The \trans{n}{v}{1238,\,1242} feature was the only case where a linear fit did not perform well because this spectral region is affected by the wing of the Lyman\,$\alpha$ absorption profile. For that line only, we modeled the continuum with a third-order polynomial to capture the more complex structure. The same continuum bandpasses were used for all three stars. 

\begin{table*}
\caption{Bandpasses Used for Continuum Normalization}
\label{tab:contnorm}
\begin{center}
\tabcolsep=0.25cm
\begin{tabular}{lcc}
Feature & Blue Continuum Bandpass & Red Continuum Bandpass  \\
 & ($\mathrm{\AA}$) & ($\mathrm{\AA}$) \\
\hline 
C\,\textsc{iii}\,1176 & $1171.0-1173.5$ & $1177.5-1181.5$ \\
N\,\textsc{iii}\,1184 & $1179.8-1181.2$ & $1186.5-1187.2$ \\
N\,\textsc{v}\,1238, 1242 & $1229.0-1232.7$ & $1248.5-1256.5$ \\
C\,\textsc{iii}\,1247 & $1245.0-1246.4$ & $1248.0-1249.0$ \\
O\,\textsc{iv}\,1341 & $1328.5-1332.0$ & $1345.0-1350.0$ \\
Fe\,\textsc{v}/C\,\textsc{iii}\,1428 & $1423.0-1425.0$ & $1433.3-1434.8$ \\
Fe\,\textsc{v}\,1445 & $1433.3-1434.7$ & $1457.0-1458.3$ \\
S\,\textsc{v}\,1502 & $1497.0-1500.0$ & $1505.0-1508.0$ \\
C\,\textsc{iv}\,1548, 1550 & $1535.0-1536.3$ & $1565.0-1565.5$ \\
Fe\,\textsc{iv}\,1570 & $1558.0-1559.5$ & $1580.5-1582.3$ \\
He\,\textsc{ii}\,1640 & $1633.0-1636.5$ & $1644.5-1646.5$ \\
N\,\textsc{iv}\,1718 & $1709.0-1714.0$ & $1725.0-1729.0$ \\
\end{tabular}
\end{center}
\end{table*}

\end{document}